\documentclass[aps,prb,twocolumn,superscriptaddress,showpacs,preprintnumbers]{revtex4}
\usepackage{tikz}
\usepackage{physics}
\usepackage[colorlinks,bookmarks=false,citecolor=blue,linkcolor=red,urlcolor=blue]{hyperref}
\usepackage{colortbl,amsthm,txfonts}
\usepackage{verbatim}
\usepackage{graphicx}
\usepackage{epsfig}
\usepackage{dcolumn}
\usepackage{bm}

\usepackage{amsmath,amssymb}

\makeatletter
\renewcommand*\env@matrix[1][*\c@MaxMatrixCols c]{%
  \hskip -\arraycolsep
  \let\@ifnextchar\new@ifnextchar
  \array{#1}}
\makeatother

\usepackage{appendix}

\begin{document}

\title{Strain induced superconductor-insulator transition on Lieb lattice}
\author{Nyayabanta Swain}
\email{nyayabanta@ntu.edu.sg}
\affiliation{School of Physical and Mathematical Sciences, Nanyang Technological University, 
21 Nanyang Link, Singapore 637371.}
\author{Madhuparna Karmakar}
\email{madhuparna.k@gmail.com}
\affiliation{Department of Physics, Indian Institute of technology, Madras, Chennai-600036, 
India.}

\begin{abstract}

We report the numerical investigation of strain induced superconductor-insulator quantum phase 
transition on a Lieb lattice. Based on a non perturbative Monte Carlo technique,  
which retains the spatial fluctuations of the superconducting pairing field at
all orders but neglects the temporal fluctuations, we show that in 
two dimensions an $s$-wave superconductor undergoes transition to a highly correlated bosonic 
insulator under the influence of strain, applied as staggered hopping amplitudes. 
We further demonstrate a strain induced BCS-BEC like crossover in the 
superconducting state, such that the superconductor-insulator transition takes place between 
a bosonic superconductor and a bosonic insulator. Our results suggest that it is the contribution 
of the dispersive bands towards the superconducting order, which dictates this crossover.
To the best of our knowledge, this is the first work to report theoretical investigation of 
``disorder free'' superconductor-insulator phase transition in systems with Lieb lattice structure.
With the recent experimental realization of the Lieb lattice in ultracold atomic gases, photonic 
lattices as well as in solid state systems, we believe that the results presented in this paper 
would be of importance to initiate experimental investigation of such novel quantum phase 
transitions. We further discuss the fate of such systems at finite temperature, 
highlighting the effect of fluctuations on the superconducting pair formations, 
thermal scales and quasiparticle behavior.
Our non perturbative numerical approach to the problem enables us to capture the 
thermal scales of the system accurately and provides us with mean field estimates of the ground 
state properties.
The high temperature quasiparticle signatures discussed in this paper are expected to serve as 
benchmarks for experiments such as radio frequency and momentum resolved radio frequency spectroscopy 
measurements carried out on systems such as ultracold atomic gases. 
 
\end{abstract}

\date{\today}
\maketitle

\section{Introduction}

Tuning the quantum behavior of a material by applying force via external strain 
has been in the forefront of research in condensed matter systems, 
over the past few years \cite{lu_pnas2017,pan_nano2019,driscoll_scadv2019,holzapfel_natcomm2013,lin_npj2019,shen_acsnano2008,hone2008}. 
A novel way of altering the lattice structure has always been 
chemical doping which exerts chemical pressure on the lattice. The technique however often 
demands for stringent experimental conditions. An alternate way to manipulate the lattice 
geometry of materials is via strain engineering. Extensive experimental works, fuelled by 
the need of designing quantum devices and materials, carried out on strain engineering over the past few 
years have shown that the technique is capable of giving rise to exotic quantum phases 
and phase transitions \cite{terrones_rev2017,hone_rev2014,amorim_rev}. 

It has been demonstrated recently that by applying tensile strain to LaCoO$_{3}$ films 
a strain induced high temperature ferromagnetic insulator could be realized which 
opens up possibility of future devices with high operation temperatures \cite{lu_pnas2017}.
Strain induced competition between charge order and interfacial superconductivity 
(with T$_{c}$ as high as $\sim 8.3$K) has been observed in SnSe$_{2}$ films grown 
on SrTiO$_{3}$ substrates \cite{pan_nano2019}.
This observation suggests the possibility of engineering two-dimensional
 (2D) materials in which strong strain and charge injection can enhance 
or even induce superconductivity \cite{pan_nano2019}. Furthermore, strain has been 
found to enhance the superconducting T$_{c}$ upto a factor of two in SrTiO$_{3}$ films 
\cite{driscoll_scadv2019}.
In the same spirit,  strain induced superconductivity with T$_{c} \sim 10$K has been 
reported for BaFe$_{2}$As$_{2}$ \cite{holzapfel_natcomm2013}. Recently, anisotropic 
in-plane strains were applied to oxygen octahedral sublattice of VO$_{2}$ and an 
intriguing behavior of in-plane orientation dependent metal-insulator transition 
was reported \cite{lin_npj2019}. Lastly, strain engineering of graphene is now an 
well established area of research with exciting promises \cite{shen_acsnano2008,hone2008,geim2009,champagne2019}.
The cumulative outcome of these experimental observations put forward strain as a new
and promising tuning parameter to control quantum many body properties. 

\begin{figure}[b]
\includegraphics[height=6cm,width=8cm,angle=0]{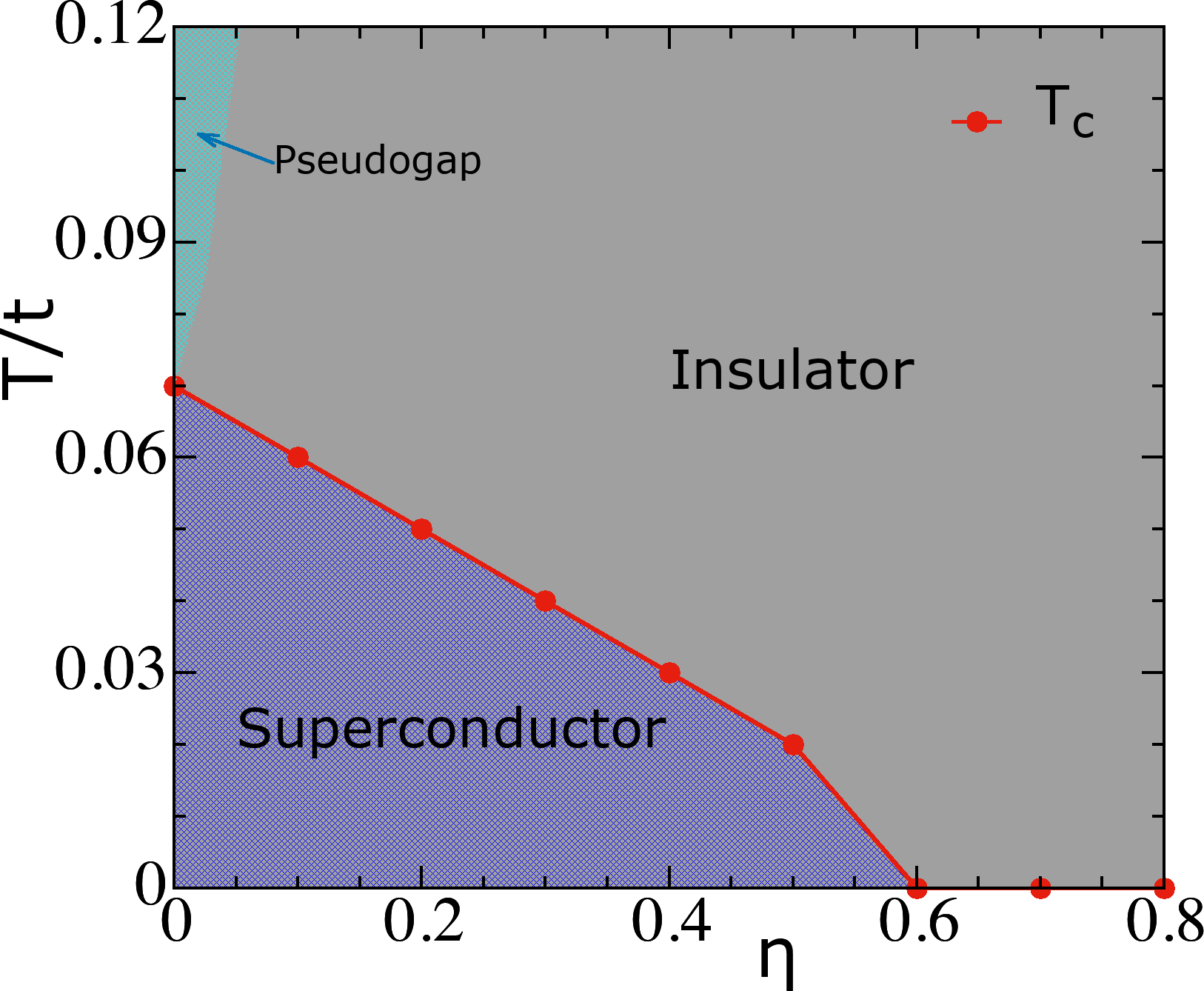}
\caption{\label{fig1} Color online: Strain-temperature ($\eta$-T) phase diagram showing the 
superconductor-insulator transition (SIT) in our model. 
The quantum phase transition occurs at a critical strain of $\eta_{c}=0.6$ for $U=2t$.  
The various phases are, (i) a gapped superconductor, (ii) a bosonic insulator, and 
(iii) a pseudogap phase. The red (dotted) curve corresponds to the T$_{c}$ of the system.}
\end{figure}

Yet another forerunner of modern condensed matter physics are the ``designer lattices'' 
\cite{flach2018}.
These are artificial lattice structures which can be engineered in ultracold atomic gases, 
photonic crystals or even in solid state systems. The lattice parameters can be tuned 
via external agencies so as to achieve the desired effect. One such category of 
designer lattices are the flat band lattices, characterized by one or more dispersionless 
spectral bands. The kinetic energy in such bands are quenched and the single particle 
spectra is independent of momentum, giving rise to the flat bands \cite{maksymenko2015,sondhi2013,liu2013}. 
While the theoretical possibility of lattices with flat bands was known since three decades \cite{sutherland1986,lieb1989,mielke1991,tasaki1992,aoki1993}, it was only 
recently that they received the renewed and much deserved interest, owing to the 
proposals of optical lattice experiments to realize designer lattices.     
One of the simplest flat band lattice in two dimension is a Lieb lattice, basically a 
depleted square lattice with three sites unit cell. The dispersion spectra 
of the same comprises of three spectral bands with two of them being dispersive
and one flat, centered at the Fermi level. The non interacting band structure consists 
of a single Dirac cone at the corners of the first Brillouin zone, intersected 
by the flat band \cite{franz2010}. 

Over the past couple of years Lieb lattice has been successfully engineered 
under different experimental settings viz. optical Lieb lattice in bosonic 
cold atoms \cite{takahashi2015,takahashi2017},  designer two-dimensional materials in which 
artificial lattices are engineered through lithography and atomic manipulations 
\cite{drost2017,swart2017,zhang2016}, optically induced photonic Lieb lattices 
\cite{molina2015,chen2016,thomson2015} etc. These experimental realizations of 
the Lieb lattice has indeed opened up a Pandora's box of future possibilities both in terms 
of experimental and theoretical investigations.   

We take cue from the two experimental advancements discussed above, viz. 
(i) strain engineering and (ii) realization of designer lattices, and attempt 
to understand the behavior of a quantum many body system on a flat band Lieb 
lattice, being subjected to strain. It is well known that interaction between 
quantum particles in flat bands can lead to spontaneous symmetry breaking and 
emergence of correlated quantum states \cite{zhou2017,shen2012,aoki2013,hu2015,chien2016,lang2017}.  
The simplest model which takes into account the effect of interaction between 
the lattice fermions is the Hubbard model and over the past couple of years substantial 
effort has been invested to analyze the physics of Hubbard model on flat band Lieb 
lattice, based on the mean field theory (MFT) \cite{torma2017,torma2017_prl,dias2015,dias2016} 
as well as other numerically sophisticated techniques such as, dynamical mean field theory (DMFT) 
\cite{torma2017_repulsive,yamashita2015,craco2018}, 
determinant quantum Monte Carlo (DQMC) etc \cite{lang2017,scalletar2016}. 
Based on the DMFT and DQMC studies, the fate of the repulsive Hubbard model on the
Lieb lattice is now relatively well established both at the ground state and at finite 
temperature \cite{torma2017_repulsive,scalletar2016}. 
Regarding the attractive Hubbard model, theoretical investigations are being carried out 
to capture the behavior of a superconducting state on the Lieb lattice.
Recently MFT has been used to map out the ground state phase diagram of 
population imbalanced fermionic superfluid on the Lieb lattice \cite{torma2017}.
Furthermore, using MFT and DMFT calculations, the attractive Hubbard model on a Lieb
lattice with staggered hopping has been studied in detail so as to understand the 
contribution of the flat band to superfluid weight in this lattice \cite{torma_natcom,torma2017_prl}.

Attempts made to access the finite temperature physics of such a system 
within the purview of MFT, would understandably overestimate the thermal scales. 
On the other hand, DMFT though gives a better estimate of the thermal scales as compared to 
the MFT, fails to capture the spatial fluctuations correctly due to its single unit cell 
approach. Away from the weak coupling limit, spatial fluctuations are expected to be 
strong and provides additional thermal scales to the system. Moreover, as we 
demonstrate in this work, interaction 
significantly renormalizes the behavior of the superconducting order even at the ground state and 
a single unit cell approach to the problem is insufficient to capture such renormalizations. 
This leaves a gap in our understanding of the physics of superconductors on a flat band Lieb lattice, 
even for an unstrained or isotropic system, and demands for a theoretical investigation to address these issues.

The interplay between strongly correlated quantum state on a flat band lattice and applied strain 
presents one with an interesting premise to understand the localization-delocalization 
transitions. It must be noted that one of the salient features of the flat band lattices 
such as the Lieb lattice is strong localization of energy, giving rise to ``compact localized 
states'' \cite{aoki1996,flach2018}. In the non interacting limit these states are immune to strain, 
and unlike the curious case of graphene \cite{shen_acsnano2008,hone2008}, strain (as applied 
through staggered hopping amplitudes) does not lead to a 
gap opening at the Fermi level. On the other hand interaction between the quantum particles 
open up a gap at the Fermi level of the flat band Lieb lattice 
and as mentioned above, gives rise to symmetry breaking quantum states, such as a 
superconducting state. The question we ask in this paper is how the application of strain 
affects this superconducting state and what are the phase transitions that it promotes?

\begin{figure}
\includegraphics[height=5.5cm,width=8cm,angle=0]{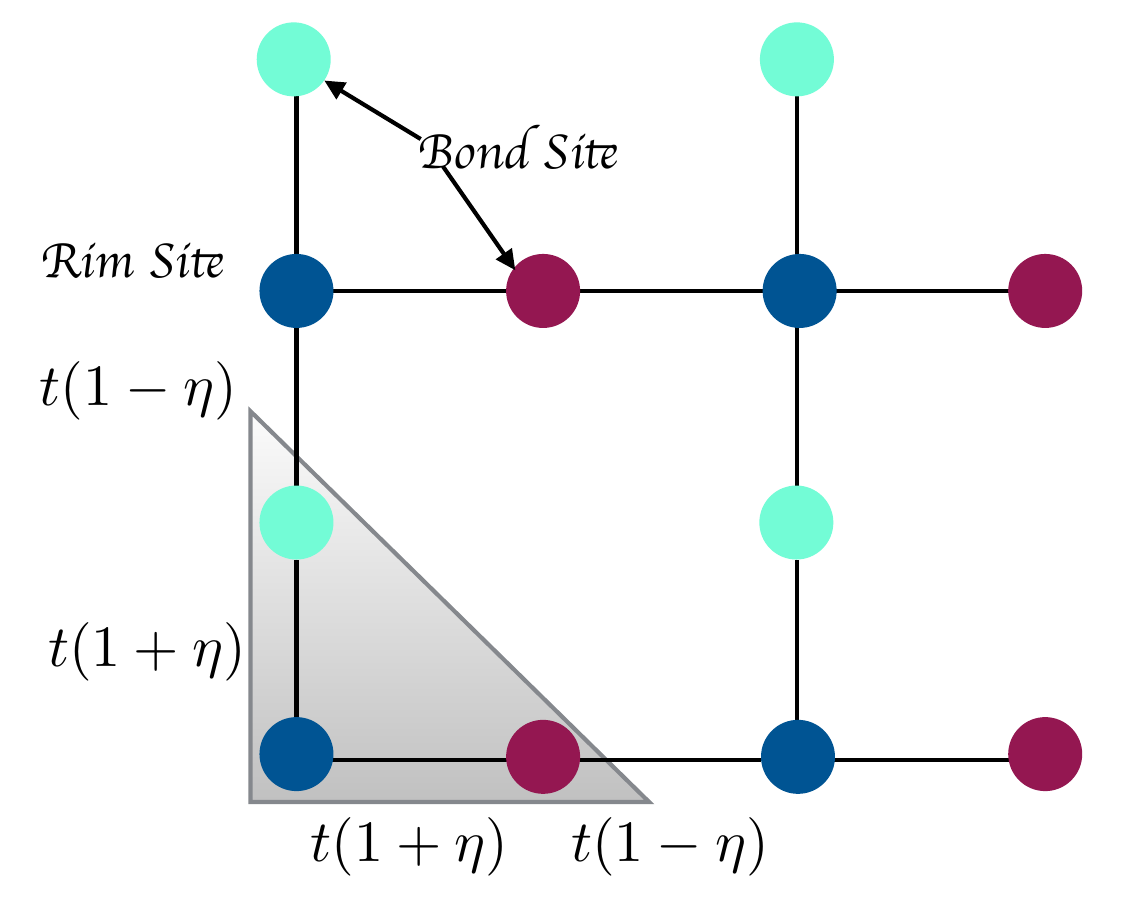}
\caption{\label{fig2} Color online: Schematic diagram showing the structure of Lieb lattice. 
The three site unit cell structure of the lattice is highlighted. 
The sites in blue are called the rim sites, while the one in red and 
green are the bond sites along the x and y-directions, respectively. 
The magnitude of strain is given by $\eta$ and it is applied as 
staggered hopping.}  
\end{figure}

Fig. \ref{fig1} constitutes the principal result of this paper where 
we demonstrate a strain ($\eta$) induced quantum phase transition between a superconductor and an 
insulator at a fixed interaction strength, on a Lieb lattice.
The transition is determined via a Monte Carlo technique which 
retains the spatial fluctuations of the superconducting pairing field at all orders but 
ignores the temporal fluctuations, thus differing from DQMC method.
However, it is complimentary to DMFT. The quantum phase transition discussed in this paper is thus
a mean field estimate of the same. While we do not expect any qualitative change in the low 
temperature results discussed in this paper, inclusion of quantum fluctuations might lead to 
quantitative changes in the estimates of the phase boundaries.
To the best of our knowledge this is the first theoretical proposal
to realize a {\it disorder free} superconductor-insulator transition (SIT) on the flat band Lieb lattice.
Consequently, our work is expected to initiate experimental investigations of the same in 
the artificial Lieb lattices, which are now an experimental reality. While we discuss our 
formalism and the results obtained from the same in the following sections we highlight 
our main observations here, (i) In the absence of strain, the system undergoes BCS-BEC 
 crossover as a function of increasing interaction $U$. The maximum T$_{c}$ is achieved 
for $U=U_{c} \sim 4$t, (ii) At a selected $U$,  strain induces a quantum SIT. 
Superconductivity is lost beyond a critical strain, through the loss of long range 
phase coherence. Notably, application of strain alters the superconducting state 
from BCS-like to BEC-like, even at weak coupling, i. e. there is a strain induced BCS-BEC crossover. 
(iii) The system is ``bosonic'' on either side of the transition, 
i. e. a BEC-like superconducting state and a bosonic insulator, respectively. The single particle 
spectral gap remains hard across the phase transition, (iv) Strain dramatically 
alters the quasiparticle spectral behavior. While in the unstrained limit the dispersion spectra 
is characterized by a single flat and two dispersive bands, the spectra at the strong strain 
limit comprises of three (gapped) flat bands, at and away from the Fermi level, where the energy 
states are strongly localized. 

The rest of the paper is organized as follows, we discuss the numerical technique used to 
carry out this work in section II, along with the superconducting and quasiparticle indicators 
based on which our results are analyzed. The results presented in section III comprises of two parts. 
In the first part we set the stage by analyzing the unstrained superconducting system on the 
Lieb lattice. The second part is dedicated to investigating the strain induced SIT at a 
particular $U$, which constitutes the focus area of this work. We discuss the relevance of our 
work from the perspective of experiments in section IV and touch upon certain aspects of the 
numerical technique used in this work. This is followed by the conclusions 
drawn based on this work.     

\begin{figure}
\includegraphics[height=6cm,width=7.5cm,angle=0]{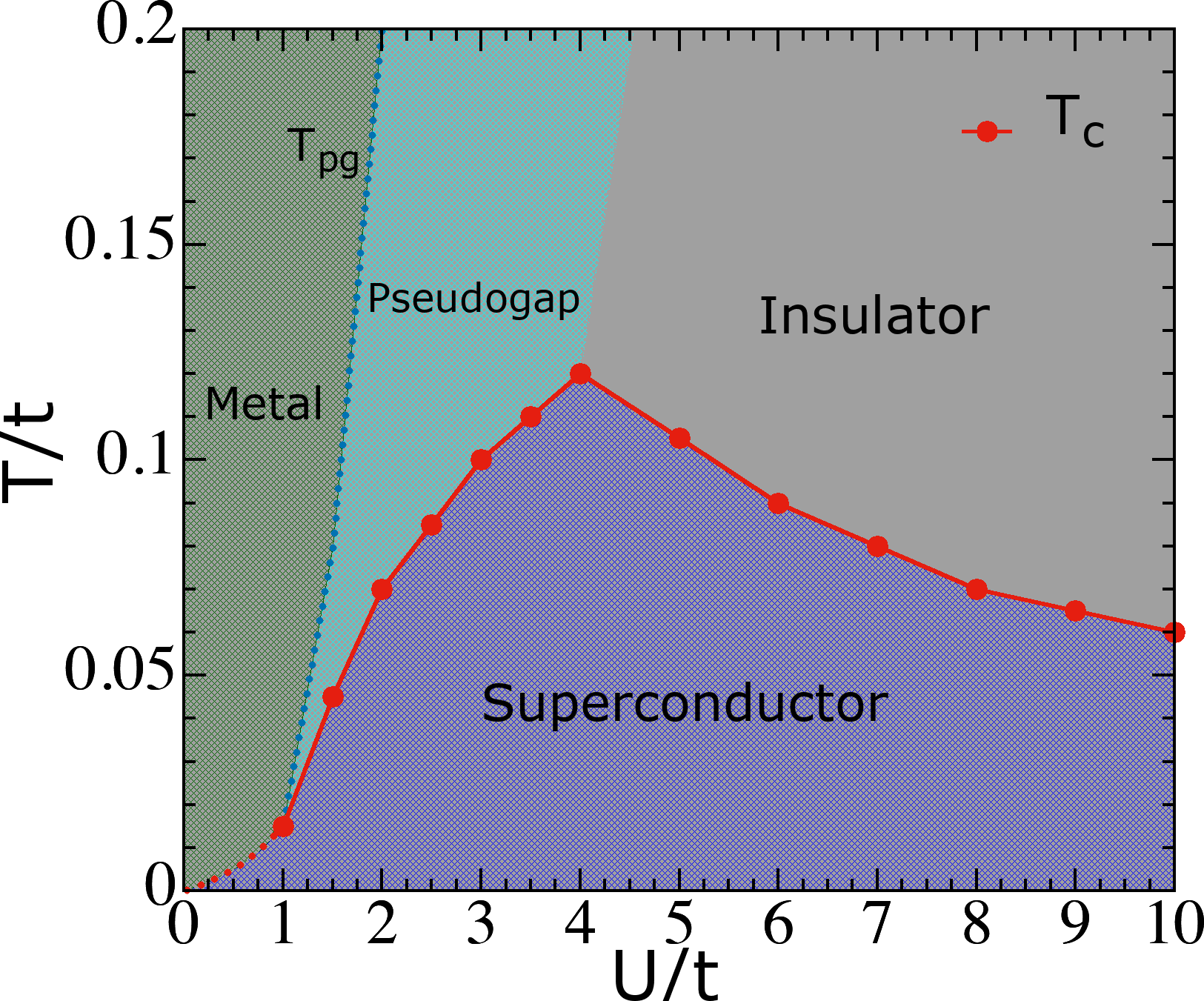}
\caption{\label{fig3} Color online: Thermal phase diagram showing the BCS-BEC crossover 
on the unstrained Lieb lattice. The behavior of T$_{c}$ is non monotonic with the peak at $U\sim 4t$. 
The ground state shows a gapped $s$-wave superconducting state. 
At the finite temperature, the weak interaction regime correspond to a metallic state, 
while the strong interaction regime is an insulator. The intermediate interaction 
regime correspond to pseudogap phase characterized by short range pair correlations.}
\end{figure}

\section{Model and method}

We define our superconducting system through a two-dimensional attractive Hubbard 
model on a Lieb lattice as \cite{torma2017_prl},
\begin{eqnarray}
H = -\sum_{\langle ij\rangle, \sigma} t_{ij}(c_{i,\sigma}^{\dagger}c_{j,\sigma} + h. c.) -
| U| \sum_{i}\hat n_{i,\uparrow}\hat n_{i,\downarrow} + \mu\sum_{i, \sigma}\hat n_{i, \sigma} \nonumber \\
\end{eqnarray}
\noindent where, $t_{ij} = (1 \pm \eta)t$ is the staggered hopping amplitude as shown in Fig. \ref{fig2}, 
for the nearest neighbors and is zero otherwise. $t=1$ sets the energy scale of the problem. 
The strain is introduced in the system in terms of staggered hopping, through the parameter $\eta$.
The unit cell of Lieb lattice comprises of three sites marked in red, blue
and green, in Fig. \ref{fig2}. In each unit cell the red and green constitutes the bond sites (being on the $x$ and $y$
bonds of a square lattice plaquette) while the blue corresponds to the rim sites (the sites corresponding to the 
square lattice plaquette). The strain ($\eta$) is introduced in a way such that increasing $\eta$ 
leads to larger hopping amplitude between the intracell sites, while simultaneously reducing the 
hopping amplitude between the neighboring unit cells.
We choose to work in a grand canonical ensemble and thus at a fixed chemical potential $\mu$;  
 $| U | > 0$ is the attractive interaction between the fermions.
 
\begin{figure*}
\centerline{
\includegraphics[height=5.5cm,width=12cm,angle=0]{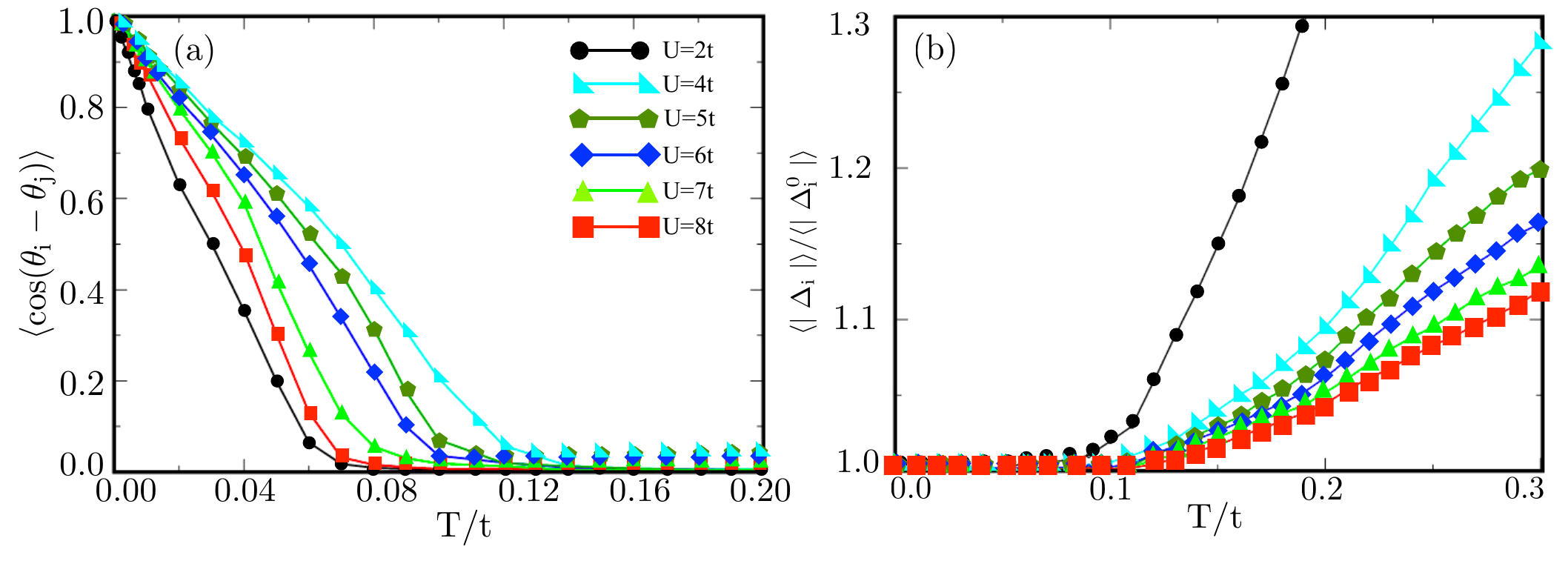}
\includegraphics[height=5.5cm,width=6cm,angle=0]{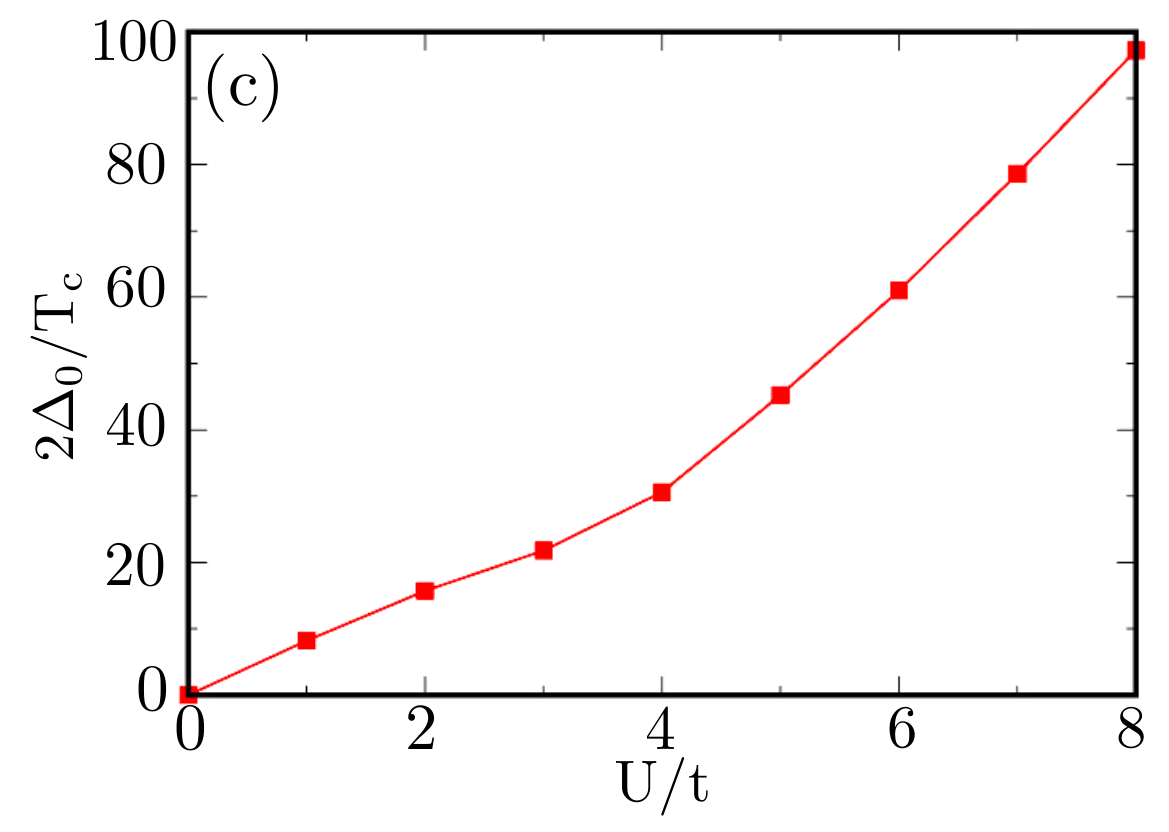}}
\caption{\label{fig4} Color online: (a) Thermal evolution of average phase correlation  
at different interactions $U/t$. The point of inflection in each curve correspond 
to the T$_{c}$. (b) Thermal evolution of the average pairing field amplitude at different 
interactions (chosen to be the same as in (a)), normalized by the corresponding value at T=0, 
in each case. (c) Mean field ratio of superconducting pairing field amplitude ($\mid \Delta_{i}^{0}\mid$) 
at T=0 and T$_{c}$. Note 
the rapid increase in this ratio with increasing $U/t$, which indicates that beyond 
the weak coupling regime the mean field theory severely over estimates the stability of 
the superconducting state.}
\end{figure*}

We decompose the four fermion term in the pairing channel using Hubbard-Stratonovich (HS) decomposition 
and introduce a ``bosonic'' auxiliary complex scalar field $\Delta_{i}(\tau) = |\Delta_{i}(\tau)| e^{i\theta_{i}(\tau)}$,  
where $|\Delta_{i}(\tau)|$ correspond to the amplitude and $\theta_{i}(\tau)$ the phase of the superconducting 
pairing field. In principle the interaction can be decomposed in charge channel as well. However, inclusion 
of a charge field along with the pairing would simply shift the Fermi level of the system without leading 
to any qualitative change in our results. We have verified that in spite of working in the grand 
canonical ensemble the number density of fermions do not drift significantly from its T=0 value 
(see appendix). 
At the same time a single channel decomposition of the interaction cuts down 
the computation cost significantly and thereby allows us to access sufficiently large system sizes, to capture 
spatial inhomogeneties. We thus restrict ourselves to the single channel decomposition and allow for two 
bosonic auxiliary fields $| \Delta_{i}(\tau)|$ and $\theta_{i}(\tau)$. 
We consider $s$-wave symmetric pairing field. 
For $| U | > $t a mean field description of the system breaks down and one needs to retain the 
fluctuations beyond the mean field. The auxiliary field comprising of both spatial and temporal 
degrees of freedom can be numerically treated exactly through DQMC, but is 
restricted to smaller system sizes (specially since the unit cell of a Lieb lattice is thrice that of an 
ordinary square lattice). 

We use an alternative Static Path approximation (SPA) \cite{evenson1970,dubi2007,tarat_epjb,karmakar2016} 
technique to address the problem wherein we retain all the spatial fluctuations (and not just the saddle point 
fluctuations) of the auxiliary fields but drop the temporal fluctuations. 
The system can thus be envisioned as free fermions moving in a random background of ``classical'' 
$\Delta_{i}$. The resulting effective Hamiltonian thus takes the form, 
\begin{eqnarray}
H_{eff} & = & -t(1\pm \eta)\sum_{\langle ij\rangle, \sigma}(c_{i, \sigma}^{\dagger}c_{j, \sigma} + h. c.) +
\mu \sum_{i, \sigma}\hat n_{i, \sigma} \nonumber \\ && + \sum_{i}(\Delta_{i}c_{i, \uparrow}^{\dagger}c_{i, \downarrow}^{\dagger}
+ \Delta_{i}^{*}c_{i, \downarrow}c_{i, \uparrow}) + \sum_{i} \frac{| \Delta_{i} |^{2}}{| U |}
\end{eqnarray}
\noindent where, the last term corresponds to the stiffness cost associated with the auxiliary field. 
This numerical technique is akin to the MFT at T=0, 
 and becomes progressively more accurate as T $\rightarrow \infty$, capturing the thermal scales accurately. 
The pairing field configurations follow the probability distribution, 
\begin{eqnarray}
P\{\Delta_{i} \} \propto Tr_{c, c^{\dagger}}e^{-\beta H_{eff}}
\end{eqnarray}
\noindent where, $\beta$ is the inverse temperature. This is related to the free energy of the fermions. 
For large and random $\{\Delta_{i}\}$ the fermion trace is computed numerically, the corresponding 
configurations are generated via classical Monte Carlo technique, diagonalizing $H_{eff}$ for each 
attempted update of $\{\Delta_{i}\}$. The relevant fermionic 
correlators are then computed on the optimized configurations at different temperatures. This numerically expensive 
technique scales with the system size as \textit{O}$(N^{4})$, where $N = 3L^{2}$ is the number of lattice sites. 
The computational cost has been cut down by using traveling cluster approximation (TCA), 
wherein instead of diagonalizing the entire lattice for each attempted update, 
we diagonalize a smaller cluster centred around the update site \cite{tarat_epjb}. 
Both SPA and TCA has been extensively bench marked for several quantum many body problems 
\cite{tarat_epjb,karmakar2016,swain2017,karmakar2018,joshi2019} and the results obtained are 
found to be in excellent quantitative agreement with those obtained by DQMC. 

\begin{figure*}
\includegraphics[height=6.0cm,width=16cm,angle=0]{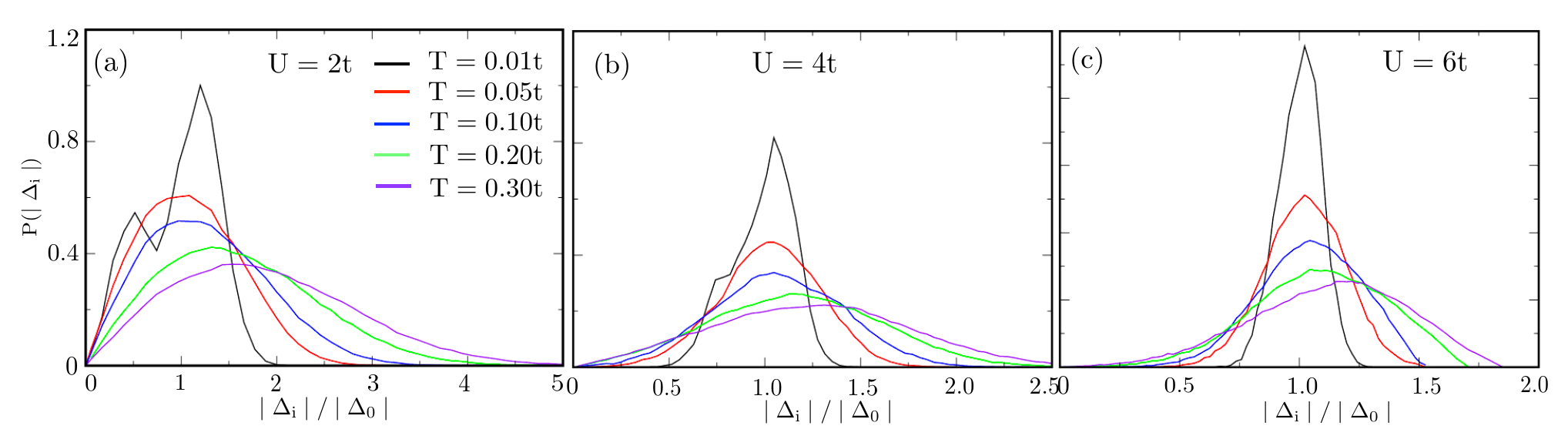}
\caption{\label{fig5} Color online: Distribution of pairing field amplitude $|\Delta_{i}|$ at different $U/t$, 
as they evolve in temperature. With increasing temperature the distribution broadens 
and the mean amplitude of $| \Delta_{i}|$ shifts towards larger values.}
\end{figure*}

The results presented in this paper corresponds to a system size of $N=768$ sites 
with the cluster size being $N_{c}=48$ sites. 
We have also verified our results with $N_{c}=108$ and have found them to be robust 
w. r. t the size of the cluster.
Finite system size analysis (upto $N=1200$) has been carried out and the results presented 
in this paper are found to be immune to system size changes.    
The parameter space encompassed by $\{\mu, | U|, \eta, T \}$ is huge and 
we restrict ourselves over relevant cross sections in this parameter space. 
We set the chemical potential to $\mu = -0.2$t (corresponding 
to a fermionic number density of $n \approx 0.9$), such that the system is
not at, but close to half filling. Owing to the $SO$(3) symmetry of the Hubbard 
model, at half filling the ground state of the system comprises of degenerate
superconducting and charge density wave orders. We have selected the filling to 
be away from this point of degeneracy such that the stable ground state is a superconductor. 
This allows us to avoid the charge density wave fluctuations and justifies our approximation 
of single channel decomposition of the Hubbard interaction.  
For the BCS-BEC crossover the interaction regime of 
t$\le | U| < 10$t has been explored, while the effect of strain is reported 
on a selected interaction cross section of $U=2$t, for $\eta \in [0:1]$. 
We have verified that our results are qualitatively immune to the choice 
of the interactions, Our ground state calculations are carried out at 
T=0.01t corresponding to one hundredth of the hopping scale and is verified of their 
robustness upto T=0.001t. 
We analyze our results based on the following superconducting and quasiparticle indicators,

\begin{itemize}
\item{Distribution of the pairing field amplitude:- \\
$P(| \Delta |) = \langle \sum_{i}\delta(| \Delta| - | \Delta_{i}| )\rangle$}
\item{Average phase correlation of pairing field:- 
$\frac{1}{N}\langle \sum_{i,j}\cos(\theta_{i}-\theta_{j})\rangle$.}
\item{Real space maps:- (a) pairing field amplitude $| \Delta_{i}|$ and 
(b) pairing field phase correlation $\cos(\theta_{0}-\theta_{i})$}
\item{Single particle density of states (DOS):- \\
$N(\omega) = \langle \frac{1}{N}\sum_{i, n}(| u_{n}^{i}|^{2}
\delta(\omega-E_{n}) + | v_{n}^{i}|^{2}\delta(\omega + E_{n}))\rangle$}
\item{Spectral function and lineshapes:- \\
$A({\bf k}, \omega) = -(1/\pi)$Im$G({\bf k}, \omega)$}
\end{itemize}
\noindent here, $G({\bf k}, \omega) = lim_{\delta \rightarrow 0}G({\bf k}, i\omega_{n})|_{i\omega_{n}\rightarrow \omega+i\delta}$, 
where,  $G({\bf k}, i\omega_{n})$ is the imaginary frequency transform of $\langle c_{\bf k}(\tau)c_{\bf k}^{\dagger}(0)\rangle$. 
$u_{n}^{i}$ and $v_{n}^{i}$ are the Bogoliubov de-Gennes (BdG) eigenvectors corresponding to the eigen values $E_{n}$ for the 
configuration under consideration.

\section{Results}

We now go back to the main observation of this work {\it i. e.} strain induced SIT, shown in Fig. \ref{fig1}. 
We note that the system undergoes a SIT at a critical strain of
$\eta_{c} \sim 0.6$ for an interaction strength of $U=2$t. 
The high temperature pseudogap regime is restricted at $\eta \rightarrow 0$, 
suggesting that strain renders the spectral gap at the Fermi level immune to thermal 
fluctuations.
Application of strain 
strongly suppresses the superconducting transition temperature (T$_{c}$) of the system.
The observation is obvious, as progressively increasing strain (as applied in our model)
decouples the three-site unit cells from each other. 

In order to analyze the different phases shown in this phase diagram,
one needs to understand the ``unstrained'' system based on the indicators
 mentioned above. We thus discuss the BCS-BEC crossover 
on the Lieb lattice in the following few sections and then focus on 
a specific interaction strength of this crossover regime to demonstrate the effect of strain, 
in the later sections of the paper. 

\begin{figure}
\includegraphics[height=10.0cm,width=8.0cm,angle=0]{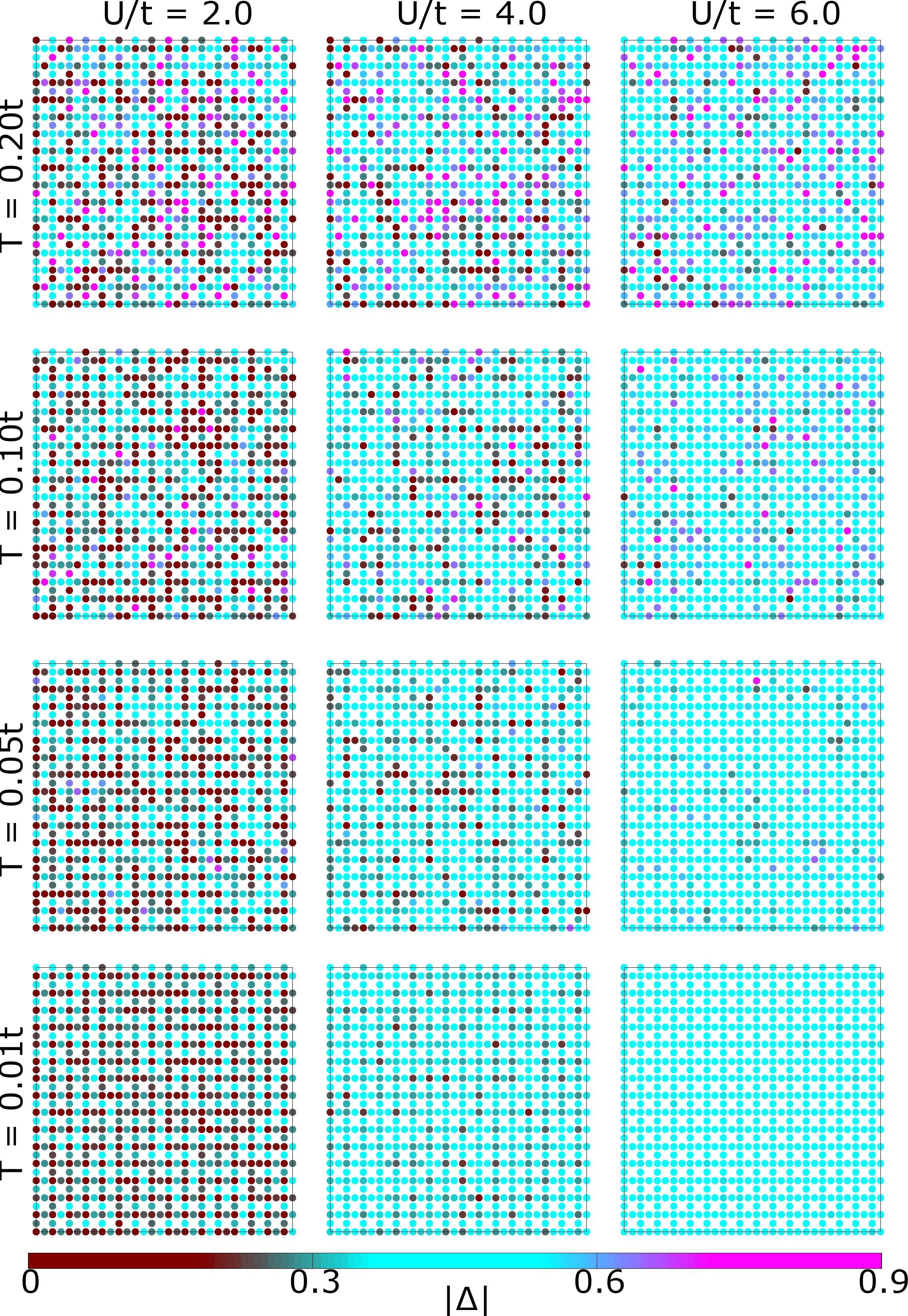}
\caption{\label{fig6} Color online: Real space maps showing the thermal evolution of the pairing 
field amplitude ($| \Delta_{i}|$) at different interaction strengths.
Note that at weak interactions and at the lowest temperature the rim and bond sites of 
the lattice have different magnitudes of $| \Delta_{i}|$, in agreement with 
the bimodal distribution of the pairing field amplitude. Each point on the spatial snapshot indicates
the pairing field amplitude at that site, with the corresponding 
weight (magnitude) given by the color bar.}
\end{figure}

\begin{figure}
\includegraphics[height=10.0cm,width=8.0cm,angle=0]{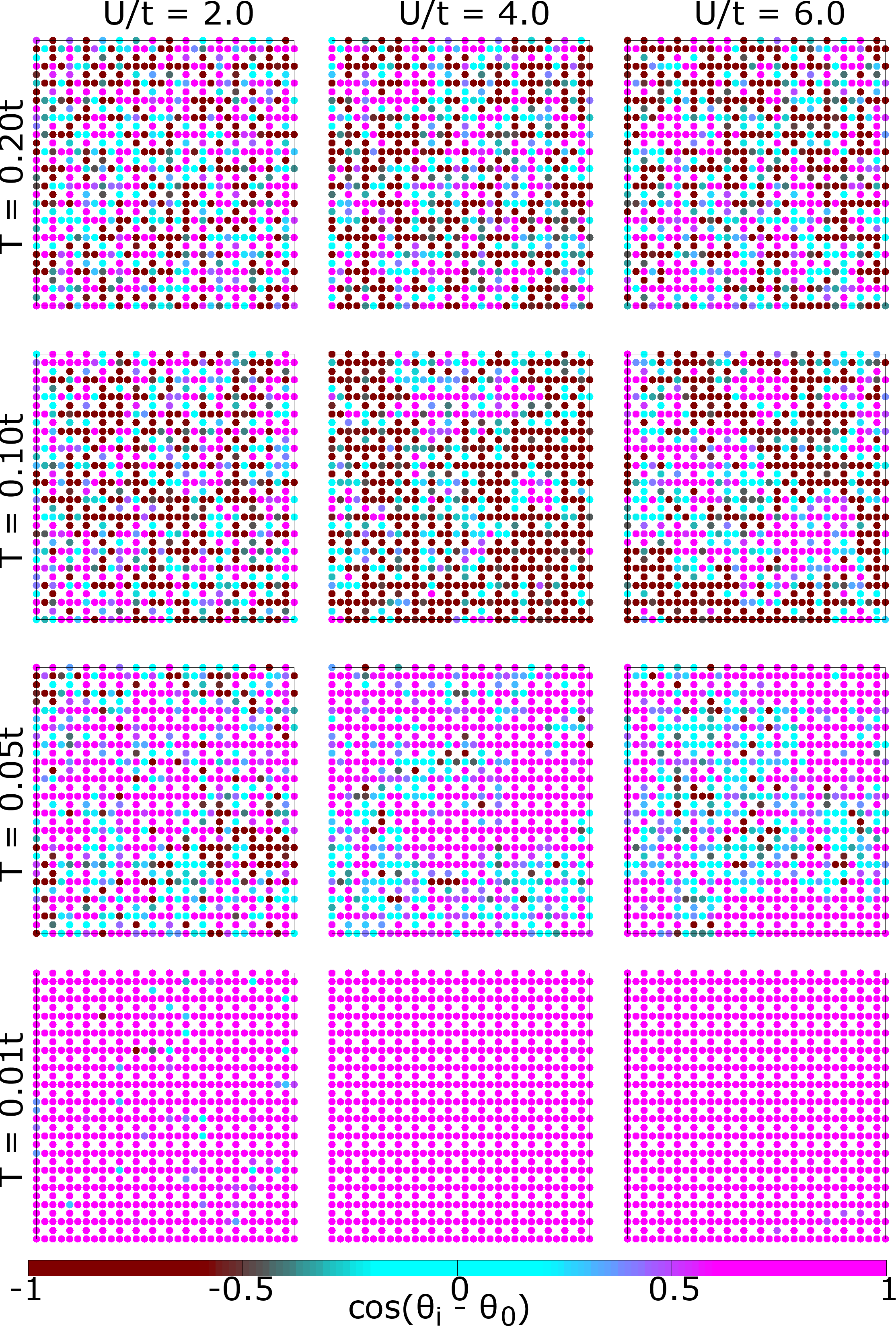}
\caption{\label{fig7} Color online: Real space maps showing the thermal evolution of the pairing
field phase correlation at different interaction strengths. While the 
low temperature state corresponds to uniform long range phase coherence at all 
interactions, the high temperature state is spatially phase uncorrelated. 
Each point on the spatial snapshot indicates the pairing field phase coherence at that lattice site, 
with the corresponding weight (magnitude) given by the color bar.}
\end{figure}

\subsection{BCS-BEC crossover}

In Fig. \ref{fig3} we present the thermal phase diagram of the BCS-BEC crossover on an unstrained 
Lieb lattice. Note that the thermal transitions discussed in this paper are 
Berezinsky-Kosterlitz-Thouless (BKT) transitions 
corresponding to the algebraic decay of long range order in two-dimensions.
The interaction-temperature ($U-T$) phase diagram shown in Fig. \ref{fig3} 
comprises of four key thermodynamic phases viz. (a) superconductor, 
(b) metal, (c) insulator, and (d) pseudogap.
The phases are demarcated by two thermal scales 
corresponding to the loss of (quasi) long range superconducting phase coherence at T$_{c}$ 
and the loss of short range superconducting pair correlations at T$_{pg}$. 
The behavior of T$_{c}$  with increasing $U$ is non monotonic, 
with the maximum (T$_{c} \sim $0.12t) at $U\sim 4$t, 
corresponding to the unitarity, in the context of ultracold atomic gases \cite{jochim2015}. 
Similar non monotonicity has been observed in the behavior of the superfluid weight as a function of 
interaction \cite{torma2017_prl}. Based on superfluid weight calculations, at half filling the 
maximum T$_{c}$ has been found to be $\sim 0.13$t at an interaction strength of $U \sim 3.5$t. The 
observation is in excellent agreement with the one presented in Fig. \ref{fig3} for the system close to half filling. 
We note that while at T$<$T$_{c}$ any finite interaction gives rise to a superconducting order, the 
high temperature T$>$T$_{c}$ phases pertaining to different interaction regimes are significantly different. 

For $U\le$t superconductivity emerges from a high temperature metallic Fermi liquid phase. 
In this regime T$_{c}$ $\sim$ T$_{pg}$, 
suggesting that the loss of short range superconducting pair correlations and 
long range phase coherence are almost simultaneous. In this weak coupling regime the loss of superconducting 
order is dictated by the suppression of pairing field amplitude and the thermal scale obeys the relation T$_{c} \sim U$. 

\begin{figure*}
\includegraphics[height=4.5cm,width=12cm,angle=0]{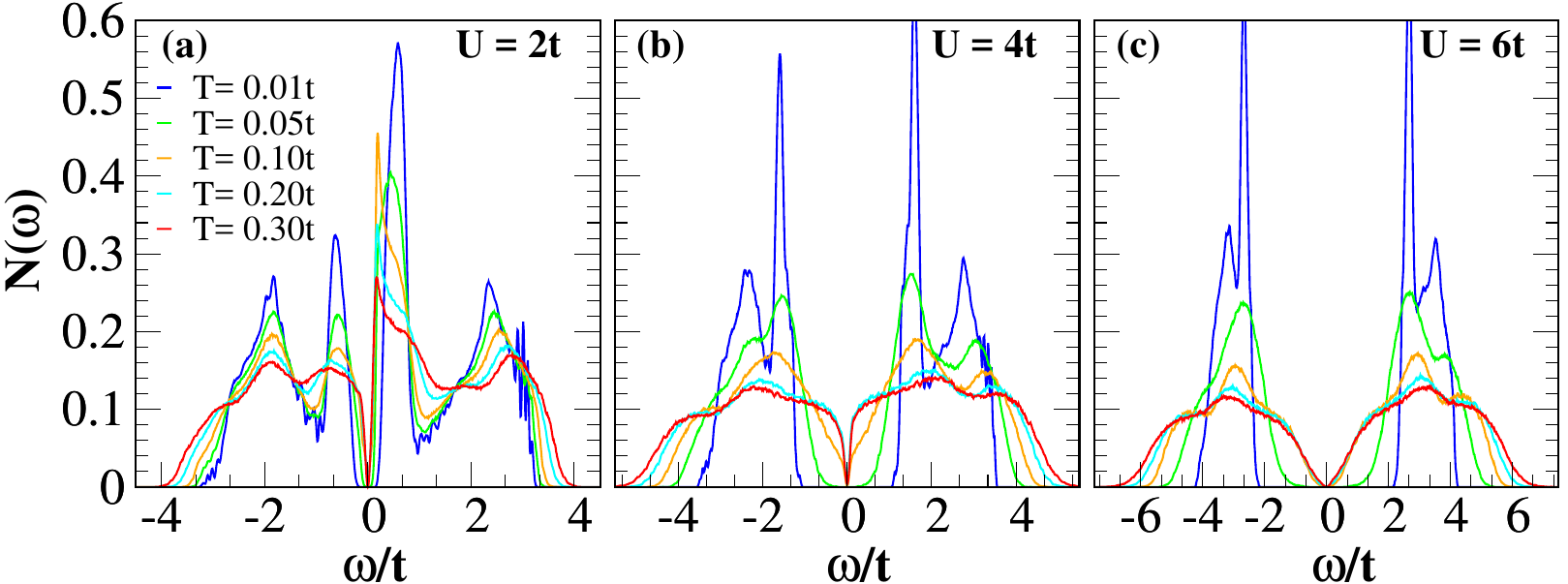}
\hspace{0.3cm}
\includegraphics[height=4.5cm,width=5cm,angle=0]{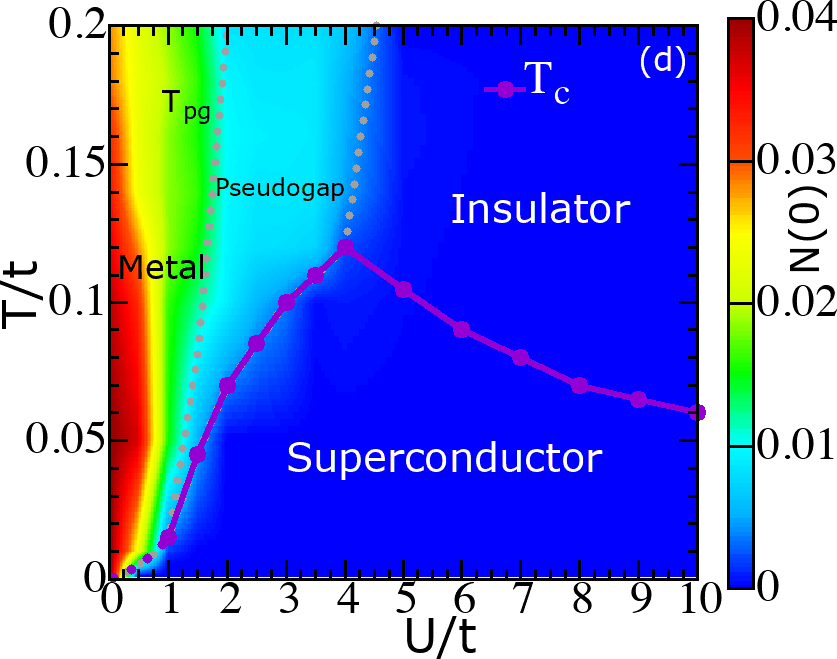}
\caption{\label{fig8} Color online: (a)-(c) Thermal evolution of the single particle density of states (DOS) at 
different interaction strengths. The loss of
superconductivity is indicated by the smearing out of the coherence peaks at high temperatures.
Thermal fluctuations pile up spectral weight at the Fermi level giving rise to pseudogap phase 
at weak and intermediate interactions.
The pseudogap regime persists for temperature 
upto T$\sim$ 1.5T$_{c}$, indicating the survival of short range pair correlations even after the long 
range order is lost. At strong interactions the DOS is gapped even at high temperatures, though 
superconductivity is already lost as suggested by the loss of coherence peaks and large transfer 
of spectral weights away from the Fermi level. The gapped phase at the high temperatures is a 
correlated bosonic insulator. (d) Map corresponding to the single particle DOS at the Fermi level, $N(0)$,
in the $U$-T plane. The color indicates the magnitude of the spectral weight.}
\end{figure*}

Over the intermediate regime of interaction t $\le U \le 4$t, the low temperature superconducting 
order emerges from a non trivial strongly correlated high temperature phase. We mark this phase as the pseudogap 
in Fig. \ref{fig3} and demonstrate that T$_{pg} >>$ T$_{c}$ in this regime, indicating the survival of 
short range pair correlations upto temperatures significantly higher than the one corresponding to the loss 
of phase coherence. The loss of superconductivity in this regime is governed by phase fluctuations and 
requires a non perturbative treatment, to get captured. Since spatial fluctuations are dominant at 
high temperatures, a single unit cell approach such as DMFT and its variants are inadequate to capture 
the behavior of the system in this regime. 

In the strong coupling regime of $U>4$t superconductivity emerges from a high temperature gapped phase 
akin to a correlated bosonic insulator. The phase is characterized by large amplitudes of the pairing field but 
a vanishing phase correlation and the thermal scale behaves as T$_{c}\sim t^{2}/U$. 
Here we note that based on DQMC calculations it was predicted that T$_{c}$ scale is strongly 
suppressed in the BEC regime on a Lieb lattice (T$_{c} \sim 0.03$t at $U=8$t), as compared to its 
square lattice counterpart \cite{scalletar2014}. Our analysis however shows a fairly robust superconducting 
order in the strong coupling regime on the Lieb lattice, with T$_{c} \sim 0.08$t at $U=8$t. We emphasize
that this discrepancy is due to strong finite size effect arising out of the small system size on which 
DQMC calculations were carried out (see appendix B).  Compared to the $N\sim 100$ used for the DQMC calculations, 
the results presented in this paper corresponds to a system size of $N=768$. 

\subsubsection{Global thermodynamic indicators}
   
Next, we show the global indicators based on which the thermodynamic phases are demarcated in 
the phase diagram. In Fig. \ref{fig4}(a) the thermal evolution of the average superconducting phase correlation 
($\langle \cos(\theta_{i}-\theta_{j})\rangle$) is presented for different $U$, where, 
$\theta_{i}$ and $\theta_{j}$ correspond to the phases of the pairing field at lattice sites 
$i$ and $j$, respectively.
The point of inflection of each curve corresponds to the T$_{c}$, at that $U$. The figure shows 
the non monotonic evolution of T$_{c}$ w. r. t $U$, with the peak T$_{c} \sim 0.12$t being at $U \sim 4$t. 
The thermal evolution of the average pairing field amplitude is shown next, in Fig. \ref{fig4}(b), normalized 
by the corresponding values at T=0. 
While the amplitude expectedly increases with $U$, the interesting observation is that 
$\langle | \Delta_{i}|\rangle \neq 0$ even when T$\gg$ T$_{c}$. The behavior is in 
remarkable contrast to the mean field theory which suggests $\langle | \Delta_{i}|\rangle = 0$
at T$\ge$ T$_{c}$. Fig. \ref{fig4}(a) and \ref{fig4}(b) together shows the impact of thermal fluctuations 
on the system, away from the weak coupling. While thermal fluctuations destroy long range 
phase coherence at T$_{c}$, short range pair correlations survive even at T$\gg$ T$_{c}$,  
leading to a non zero $\langle | \Delta_{i}| \rangle$.
Finally, Fig. \ref{fig4}(c) shows the superconducting gap vs T$_{c}$ ratio as a function of increasing 
interaction. In the BCS limit this ratio is $3.5$. As shown in Fig. \ref{fig4}(c),  at $U=2$t the ratio 
$\sim 18$, which is significantly above the BCS prediction and grows as $\sim (U/t)^{2}$ at 
large $U$. The figure demonstrates that the T=0 gap is not a suitable indicator of the robustness 
of the superconducting state, beyond the weak coupling regime. 

\begin{figure*}
\includegraphics[height=12cm,width=16cm,angle=0]{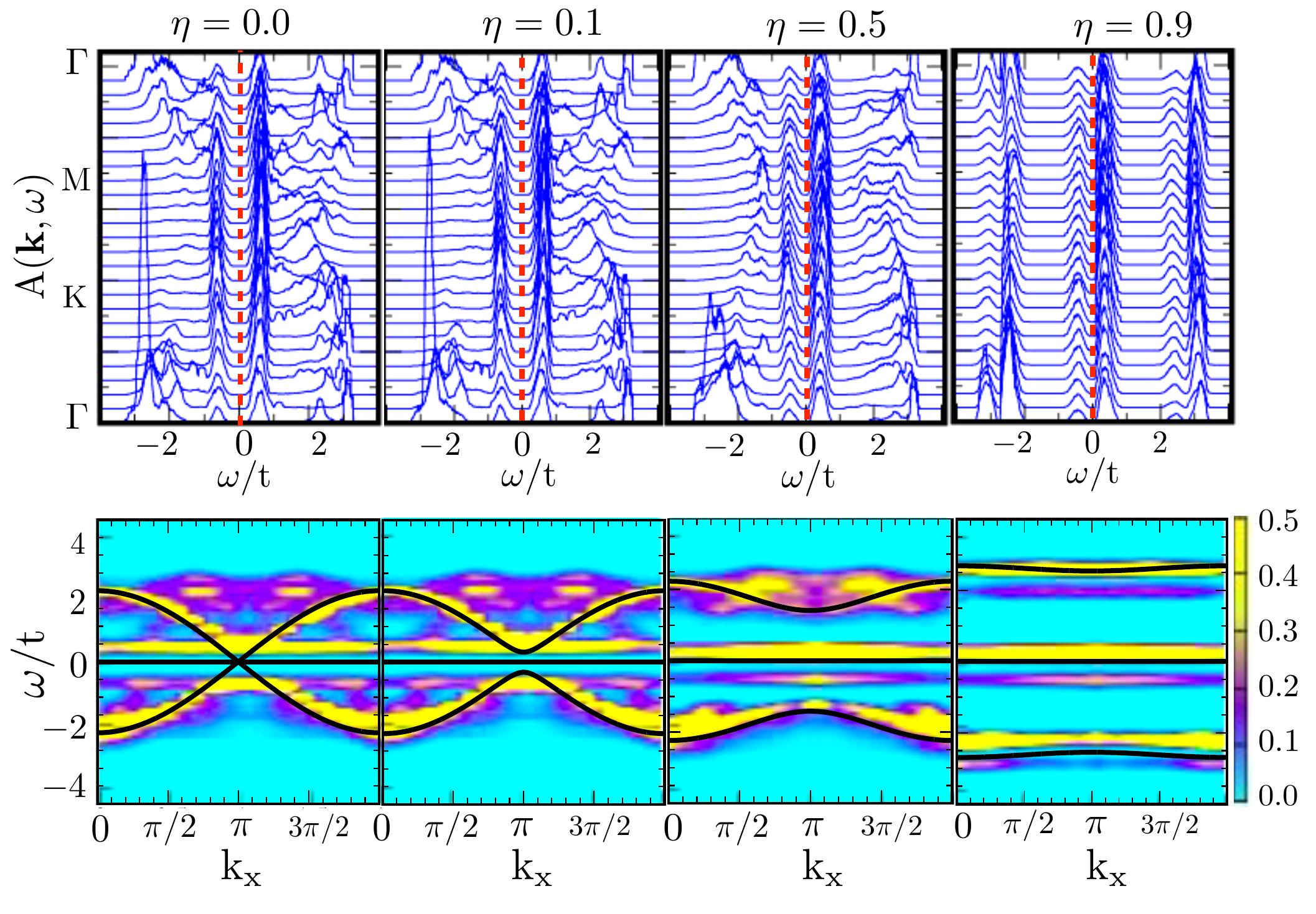}
\caption{\label{fig9} Color online: Quasiparticle signatures at T=0 and $U=2t$ as a function of increasing strain
($\eta$), expressed in terms of the (i) spectral lineshape and (ii) spectral function ($A({\bf k}, \omega)$). 
The top row shows the spectral lineshape along the trajectory $(0, 0) \rightarrow (\pi, 0) 
\rightarrow (\pi, \pi) \rightarrow (0, 0)$ across the Brillouin zone. Note that the flat 
band at the Fermi level remains immune to the effect of strain, however the dispersive bands
away from the Fermi level progressively flattens out with increasing $\eta$. In the bottom row 
the spectral function evolution along the trajectory $(q_{x}, \pi)$ is shown as the function of increasing 
strain. At large strain the spectrum comprises of localized energy bands only. The dispersion spectra 
as obtained in the non interacting limit ($\mid U\mid$=0) is shown as solid black curves on the maps.
The effect of interaction is to open up a gap at the Fermi level.}
\end{figure*}

In order to highlight the effect of thermal fluctuations on the pairing field, we next show 
the distribution of the pairing field amplitude at different temperatures for selected 
interactions representative of the (i) weak ($U = 2$t) (we select $U=2$t for the weak coupling limit 
since for $U \lesssim $t the correlation length ($\xi$) of the superconducting pairs become comparable 
to our system size), (ii) intermediate 
($U = 4$t) and (iii) strong ($U = 6$t) coupling regimes, in Fig. \ref{fig5}. 
For any interaction, at the lowest temperature the amplitude of the pairing field $| \Delta_{i}|$ exhibits 
a narrow distribution (ideally a delta function), with the mean corresponding to the T=0 
mean field value $| \Delta_{0} |$. Here we note that at $U=2$t the distribution is bimodal at the 
lowest temperature, showing that there are two different contributions to the superconducting order  
and the mean amplitude of $| \Delta_{i}|$ is different at the rim and 
at the bond sites. This behavior is specific to the bipartite nature  
of the Lieb lattice. At weak interactions, the effect of the underlying lattice is dominant 
and thus the distinction of $| \Delta_{i}|$ at the bond and rim sites show up in 
the distribution. Distinction between the local order parameters corresponding to the 
rim and bond sites have been reported in the literature. Based on DMFT calculations it 
was demonstrated that while the local order parameter corresponding to the rim and bond sites 
vanish at the same temperature T$_{c}$, 
their magnitudes are significantly different at T$\neq$T$_{c}$\cite{torma2017_prl}. The calculations were 
however restricted to the weak coupling regime of upto $U\sim 2$t. In Fig. \ref{fig5} we demonstrate 
how interaction progressively renormalizes the distribution, such that for intermediate and 
strong interactions the distribution is unimodal, indicating uniform contribution to the 
superconducting order by the bond and rim sites.  
Progressive rise in temperature broadens out the distribution and shifts the mean amplitude 
towards larger values of $| \Delta_{i}|$. Moreover, with increasing interaction the 
width of the distribution reduces, indicating the reduction in the coherence length of the 
Cooper pair. In the next section we demonstrate how this transition in distribution from bimodal to 
unimodal, bears out in the real space.  

\subsubsection{Real space maps}

Fig. \ref{fig6} and \ref{fig7} show the spatial snapshots of the pairing field amplitude 
and phase correlation for a single Monte Carlo configuration, as the system evolves in temperature, 
at different interactions. 
Each point on the spatial snapshot corresponds to the pairing field amplitude/phase correlation 
at that site on the Lieb lattice, with the weight indicated by the color coding. 
In agreement with the bimodal distribution of $| \Delta_{i}|$ discussed 
above, at $U=2$t and at the lowest temperature the map corresponding to pairing field 
amplitude shows different magnitude at the rim and bond sites. We note that the pairing 
field amplitude is large at the bond sites as compared to that of the rim sites. The 
contribution to the flat band is through the bond sites only, while the rim sites 
give rise to the dispersive bands. Our real space maps show that the flat band leads 
to a larger contribution to superconducting pairing field amplitude as compared to 
the contribution by the rim sites. The observation is in agreement with the inference 
drawn on the basis of superfluid weight in ref.\cite{torma2017_prl}, wherein the flat 
band has been shown to give a larger contribution to the superfluid weight (as geometric 
weight) as compared to the contribution by the dispersive bands, in the regime of weak 
coupling. Away from the weak coupling, we find that the contribution to the superconducting 
pairing field from the flat and dispersive bands are equal.
In agreement with the unimodal distribution of $| \Delta_{i}|$, the real space 
map is homogeneous at the lowest temperature, for intermediate and strong coupling. While 
thermal fluctuations tend to randomize the high temperature state at all interactions the 
effect is less pronounced at $U=6$t, owing to the large superconducting gap. 

The long range phase coherence is robust at the lowest 
temperature at any finite interaction. Increase in 
temperature leads to regions where the phase coherence is strongly suppressed. 
Consequently, there are regions of ``local'' superconducting phase that survive upto 
high temperatures leading to the pseudogap phase shown in the phase diagram. 
Here the system behaves as a collection of Josephson junctions without any phase coherence 
between them. In the strong coupling regime, the system loses phase coherence completely 
at T$>$ T$_{c}$ even though $| \Delta_{i}|$ continues to be large.  

\subsubsection{Quasiparticle signatures}

The contribution of the flat and dispersive bands towards superconducting pairing is next 
analyzed based on the quasiparticle signatures. In Fig. \ref{fig8}, we show the single particle 
density of states (DOS) at the selected interaction strengths of $U=2$t, $4$t and $6$t. 
The low temperature state at any interaction bears signature of $s$-wave superconducting order 
in terms of a hard gap at the Fermi level and sharp coherence peaks at the gap edges. 
The prominent satellite peaks away from the Fermi level correspond to the contributions of the spectral weight 
from the dispersive bands. 
In the weak coupling regime of $U=2$t, thermal fluctuations rapidly pile up spectral weight at the 
Fermi level and for T$\ge 0.1$t global superconductivity is lost as indicated by the smearing out of the 
coherence peaks. The finite spectral weight at the Fermi level signifies the survival of short range 
pair correlation, corresponding to the pseudogap phase. 
The characteristics of spectral weight contributions from the flat and the dispersive 
bands are very different. The spectral weight contribution of the flat band is independent of 
momentum, while the dispersive bands exhibits a momentum dependent gap minima.

\begin{figure}
\includegraphics[height=5cm,width=9cm,angle=0]{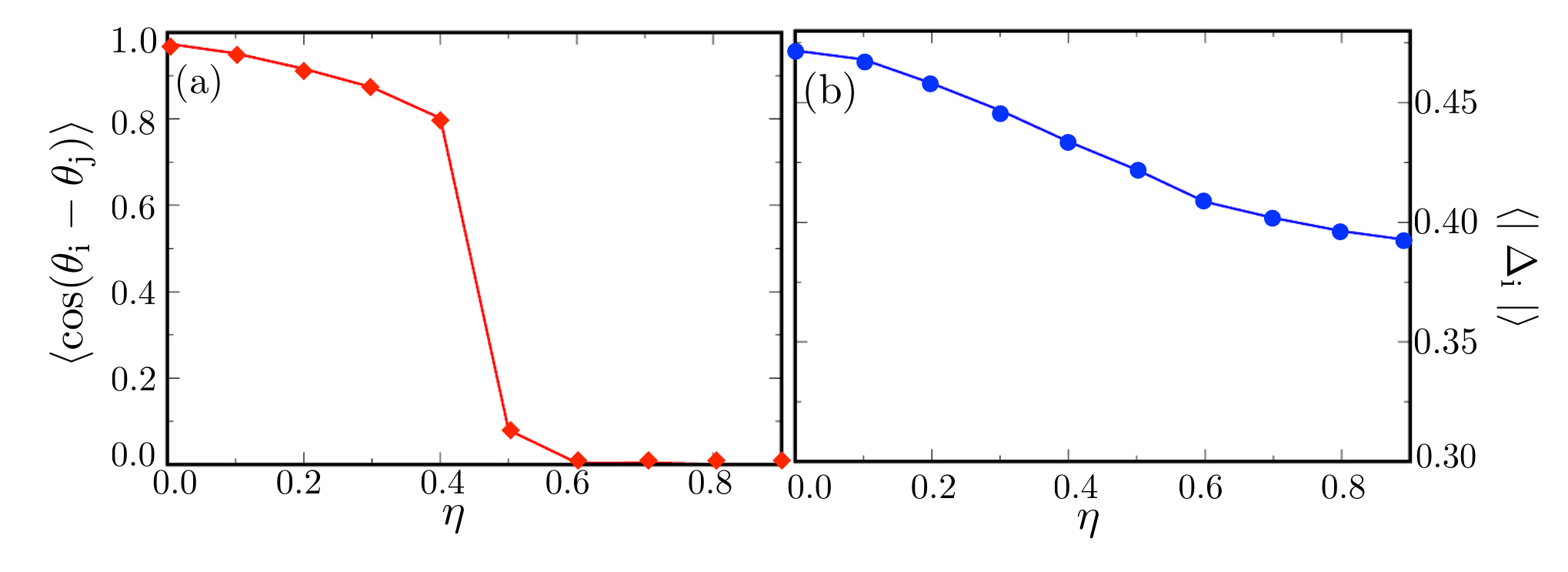}
\caption{\label{fig10} Color online: (a) Average pairing field phase correlation and (b) average of pairing field 
amplitude at T=0 as a function of increasing strain ($\eta$) at $U=2t$. The system loses 
its long range phase coherence at $\eta = \eta_{c} \sim 0.6$ as shown by (a). The pairing 
field amplitude however survives at $\eta > \eta_{c}$.}
\end{figure}

In the intermediate coupling regime ($U=4$t), interaction renormalizes the band structure and leads 
to smearing out of the dispersive bands. Consequently, the satellite peaks away from the Fermi level 
are now less prominent. Upto a temperature of T$\sim 0.05$t the superconducting gap 
at the Fermi level reduces monotonically. At still high temperatures the coherence peaks smear out 
and there is a small but finite weight at the Fermi level, over the temperature regime of  
0.12t $\le$ T $<$ 0.2t. The observation is characteristic to the survival of short range pair correlations 
in the pseudogap phase. The high temperature phase manifests strong effect of the flat band 
localization in terms of immunity of the single particle DOS at the Fermi level towards thermal fluctuations. 

In the regime of strong coupling ($U=6$t), the gap persists at the Fermi level even at high temperatures
T$\sim 0.3$t. However, unlike the superconducting gap at the low temperatures, the high temperature 
gap is non superconducting and arises out of a strongly correlated bosonic insulating state, as suggested by 
the absence of coherence peaks. We understand the origin of this high temperature gap as follows: 
the strong coupling regime is characterized by a large pairing field amplitude $| \Delta_{i}| \sim U$. 
At low temperatures this large $| \Delta_{i}|$ gives rise to a large superconducting gap, as the 
temperature increases even though the phase coherence is lost the $| \Delta_{i}|$ continues to 
be large. Over a narrow window of temperature the $| \Delta_{i}|$'s changes from being perfectly phase
correlated to randomly oriented. These randomly oriented but ``large'' $| \Delta_{i}|$'s not only opens up 
a gap at the Fermi level, but also broadens the spectra by transferring large weight away from the Fermi level. 
The contribution to the spectral weight from the dispersive bands have reduced significantly 
at this interaction as suggested by the vanishing satellite peaks. 
\begin{figure}
\includegraphics[height=8.5cm,width=8.5cm,angle=0]{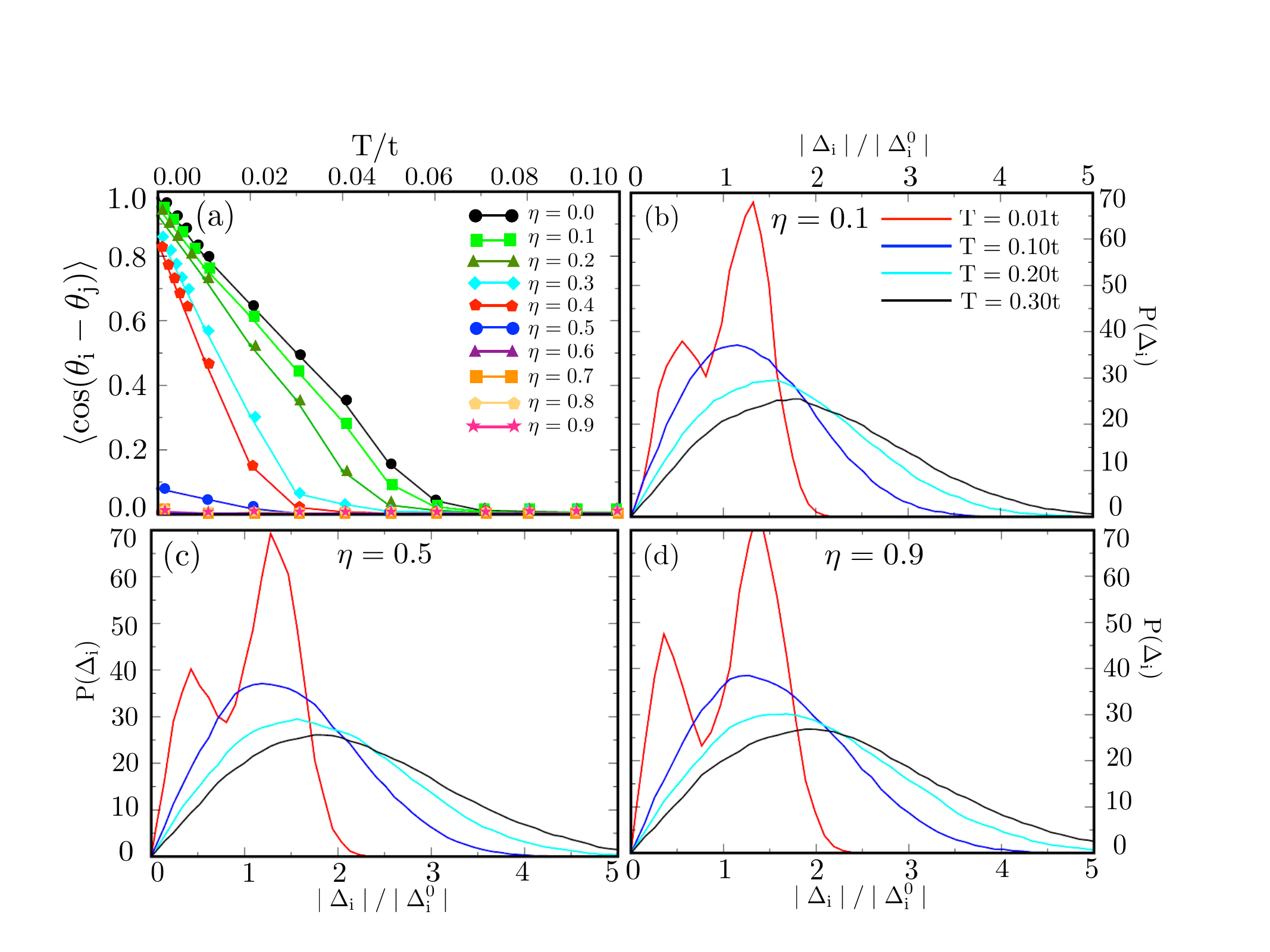}
\caption{\label{fig11} Color online: (a) Thermal evolution of average pairing field phase correlation with 
increasing strain ($\eta$) at $U=2t$. The point of inflection in each curve indicates the $T_{c}$. 
Note how phase correlation is strongly suppressed beyond $\eta\sim 0.4$ 
and is lost at $\eta_{c} \sim 0.6$. The distribution of the pairing field amplitude 
at selected strain cross sections of $\eta=0.1$, $0.5$ and $0.9$ are shown in panels 
(b), (c) and (d), respectively. At the lowest temperature the distribution is bimodal 
at all strain regimes, showing that the bond and rim sites of the underlying lattice contains 
pairing fields of different amplitudes. Thermal fluctuations smooth out the distributions 
and makes them unimodal. With increasing temperature the mean amplitude of the pairing field 
shifts towards larger values, at all strains. Note that even in the insulating regime the 
pairing field amplitude remains fairly robust.} 
\end{figure}
In Fig. \ref{fig8}(d), we highlight the quasiparticle behavior at the Fermi level 
in the $U-T$ plane, by mapping out the single particle DOS ($N(0)$). 
The T$<$T$_{c}$ regime correspond to gapped superconducting state as suggested 
by the vanishing DOS at the Fermi level. In the regime of strong interactions 
the high temperature phase is a correlated bosonic insulator,
and thus correspond to a gapped quasiparticle spectrum. 
The weak interaction regime is characterized by large spectral weight at the Fermi level, 
corresponding to the metallic phase at high temperature. 
The pseudogap phase in the intermediate interaction regime is characterized by 
a small finite spectral weight and a prominent dip, at the Fermi level.

\subsection{Strain induced superconductor-insulator transition}

In the previous section we have established the behavior of an $s$-wave superconductor  
on an unstrained Lieb lattice. Next we focus on the principal aspect of this work, wherein we 
subject this $s$-wave superconductor to strain,  applied through staggered hopping. We have shown 
that the loss of superconductivity at high temperature is dictated purely by the loss of phase coherence,  
except at the weak coupling regime, where the loss of superconductivity 
is due to the pairing field amplitude fluctuations. We now ask the following questions: 
what is the fate of this system when the underlying lattice is deformed by applied strain? 
Whether superconductivity survives in presence of the applied strain or is there a critical 
strain beyond which the superconducting order is lost? 
What is the nature of the phases across such a transition? We attempt to answer these questions
in the present section.
 
We select a particular interaction strength of $U=2$t, such that the effect of the 
underlying lattice structure is not smeared out by strong interactions, 
in the BCS-BEC crossover picture and tune the magnitude of 
applied strain through $\eta$ (see Fig. \ref{fig2}). In the absence of strain ($\eta=0$) the 
system at this interaction evolves from a gapped superconducting ground state to a 
pseudogapped high temperature phase, with increasing temperature.
Straining leads to reconstruction of the band structure 
of the lattice. We begin our analysis by characterizing this reconstruction of the 
band structure in terms of (i) spectral line shapes and (ii) spectral function, 
at the ground state. 

\subsubsection{Band structure reconstruction}

Fig. \ref{fig9} shows the strain dependent evolution of the momentum resolved spectral 
function $A({\bf k}, \omega)$ at $U=2$t. The top panels of Fig. \ref{fig9} shows the spectral 
lineshape along the trajectory $(0, 0) \rightarrow (\pi, 0) \rightarrow (\pi, \pi) 
\rightarrow (0, 0)$ across the Brillouin zone, at selected strain values. 
The flat band ensures a momentum independent gap at the Fermi level across the momentum trajectory 
mentioned above. Away from the Fermi level the spectra shows two dispersive bands, which 
are significantly less robust as compared to the flat band. Increasing strain progressively 
weakens the coupling between the unit cells, and as a consequence 
the spectral weight contribution from the dispersive bands reduce. At the same time the 
intracell coupling increases with strain, leading the flat band at the Fermi level to 
renormalize the dispersive bands and flatten them out with 
increasing strain ($\eta=0.5$ and $0.9$). The long range superconducting order gets progressively 
destroyed as the unit cells decouple from 
each other. The large strain regime is thus an insulator with three flat bands, each split into two. 

In a Lieb lattice the dispersive bands touch the flat band at the M-point 
($\pi, \pi$) in the Brillouin zone. Consequently, the effect of strain is most pronounced along the 
trajectory containing the M-point. We show the spectral function maps along the ($k_{x}, \pi$) 
trajectory as they evolve with strain, in the bottom row of Fig. \ref{fig9}. Further, we 
compare the dispersion spectra with that obtained at the non interacting limit (black curves)
so as to highlight the effect of interactions on the system.

In the non interacting and unstrained limit ($\eta$=0) the dispersion spectra 
is analytically tractable and is given as,

\begin{eqnarray}
E_{\pm {\bf k}}=\pm 2t\sqrt{1+(\cos k_{x}a+\cos k_{y}a)/2}
\end{eqnarray}

Application of strain progressively
pushes the dispersive bands away from the flat band and 
flattens them out. The corresponding dispersion relation reads as,

\begin{eqnarray}
E_{\pm {\bf k}} = \pm 2t \sqrt{1+\eta^{2}+(1-\eta^{2})(\cos k_{x}a+\cos k_{y}a)/2}
\end{eqnarray}

\noindent where,  
$a$ is the lattice spacing \cite{torma2017_prl}. The flat band by itself is immune to strain and unlike graphene 
does not undergo a gap opening just by the application of strain. Interaction opens up a gap 
at the Fermi level irrespective of the 
magnitude of the strain and choice of the trajectory across the Brillouin zone. As the strain 
is increased and the dispersive bands are pushed away from the flat band, they split into two 
as they flatten out progressively with strain. In the limit of large 
strain ($\eta=0.9$) the spectra comprises of three flat bands (each split into two) with localized energy. 
The dispersion spectra carries crucial information regarding the momentum dependence of the 
underlying superconducting state.

As mentioned before there are two separate contributions to the superconducting 
order, arising out of the flat and the dispersive bands. A momentum dependent spectra with a finite momentum gap 
minima is a signature of BCS-like superconducting state \cite{trivedi2016,randeria_taylor}. While on the other hand 
a spectral gap at ${\bf k}=0$ corresponds to a superconducting state akin to BEC.  
As the name suggests,  the spectral weight contribution of the  
flat band is always independent of momentum. In the superconducting regime ($\eta<\eta_{c}$) the 
contribution of the dispersive bands undergo a crossover akin to the BCS-BEC, as a function of 
increasing strain, such that the pairing field amplitude remains robust against the applied 
strain but the phase coherence is progressively lost. Based on the inferences we draw from the 
strain induced band structure reconstruction, we suggest that close to the SIT the scenario 
is dominated by the spectral weight contribution from the flat band. It is tempting to call this 
contribution as ``bosonic'', owing to the lack of finite momentum dependence of the spectra. 
In that spirit, the spectral weight contribution from the flat band is always bosonic, even in 
the unstrained limit. In the limit of large interaction (at $\eta=0$) the dispersive bands merge
with the flat band giving rise to a momentum independent dispersion spectra, and in turn a 
bosonic contribution. 
The crossover from a BCS like physics to a BEC like state w. r. t either 
$\eta$ or $U$ is always determined by the contribution of the dispersive bands. It must however 
be kept in mind that the indicators discussed in this work are based on,  
(i) single particle correlations and (ii) MFT like approach towards quantum phase transition.
Whether a more sophisticated approach to this problem via DQMC, characterized through two particle 
correlations modify this picture is an intriguing aspect worth investigating, but is out 
of the scope of the present work.
\begin{figure*}
\includegraphics[height=9cm,width=16cm,angle=0]{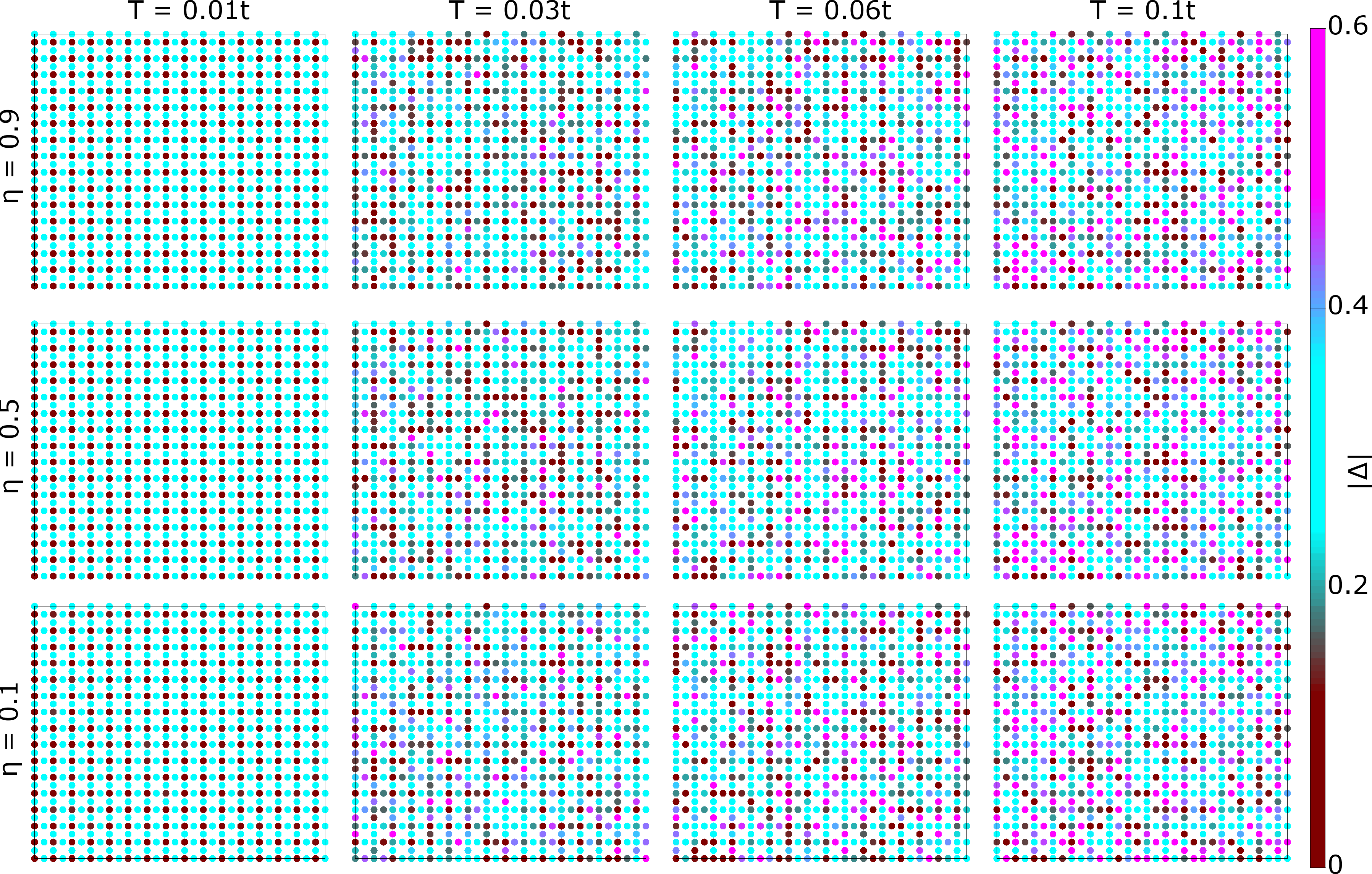}
\caption{\label{fig12} Color online: Real space snapshots of pairing field amplitude as it evolve with 
temperature at $U=2t$ and selected strain cross sections. The lowest temperature phase shows 
bimodal distribution of $| \Delta_{i}|$ at all $\eta$'s. Each point in the map indicates
the pairing field amplitude at that particular lattice site with the corresponding weight 
(magnitude) being given by the color bar.}
\end{figure*}
 
\subsubsection{Global indicators}

We next show the global indicators in terms of average phase correlation and average 
pairing field amplitude as a function of strain at the ground state,  in Fig. \ref{fig10}. We note that while 
the long range phase coherence is destroyed by strain, the pairing field amplitude is
only weakly suppressed. This is a significant observation, which shows that even at the 
ground state it is the loss of long range phase coherence that kills off the superconducting 
order. Recent DMFT calculations have suggested that the loss of superconductivity at a critical 
strain is dictated by the collapse of local superconducting order parameter \cite{torma2017_prl}. 
We argue that the critical strain should be 
determined based on the loss of long range phase coherence and that the pairing field 
amplitude is always finite. It would be interesting to 
probe the superconducting order locally for intermediate regime of strain (close to $\eta_{c}$) 
where even though the pairing field amplitude is robust, the (quasi) long range phase coherence 
is strongly suppressed. This can be achieved via scanning tunnelling microscopy (STM) and 
tunnelling conductance measurements which can probe the spatial order locally. One would 
expect a progressive disappearance of the coherence peaks in the local density of states, 
with increasing strain.
Fig. \ref{fig10} shows that at $U=2$t the system undergoes a quantum superconductor-insulator 
transition at the critical strain of $\eta_{c} \sim 0.6$. 

\begin{figure*}
\includegraphics[height=9cm,width=16cm,angle=0]{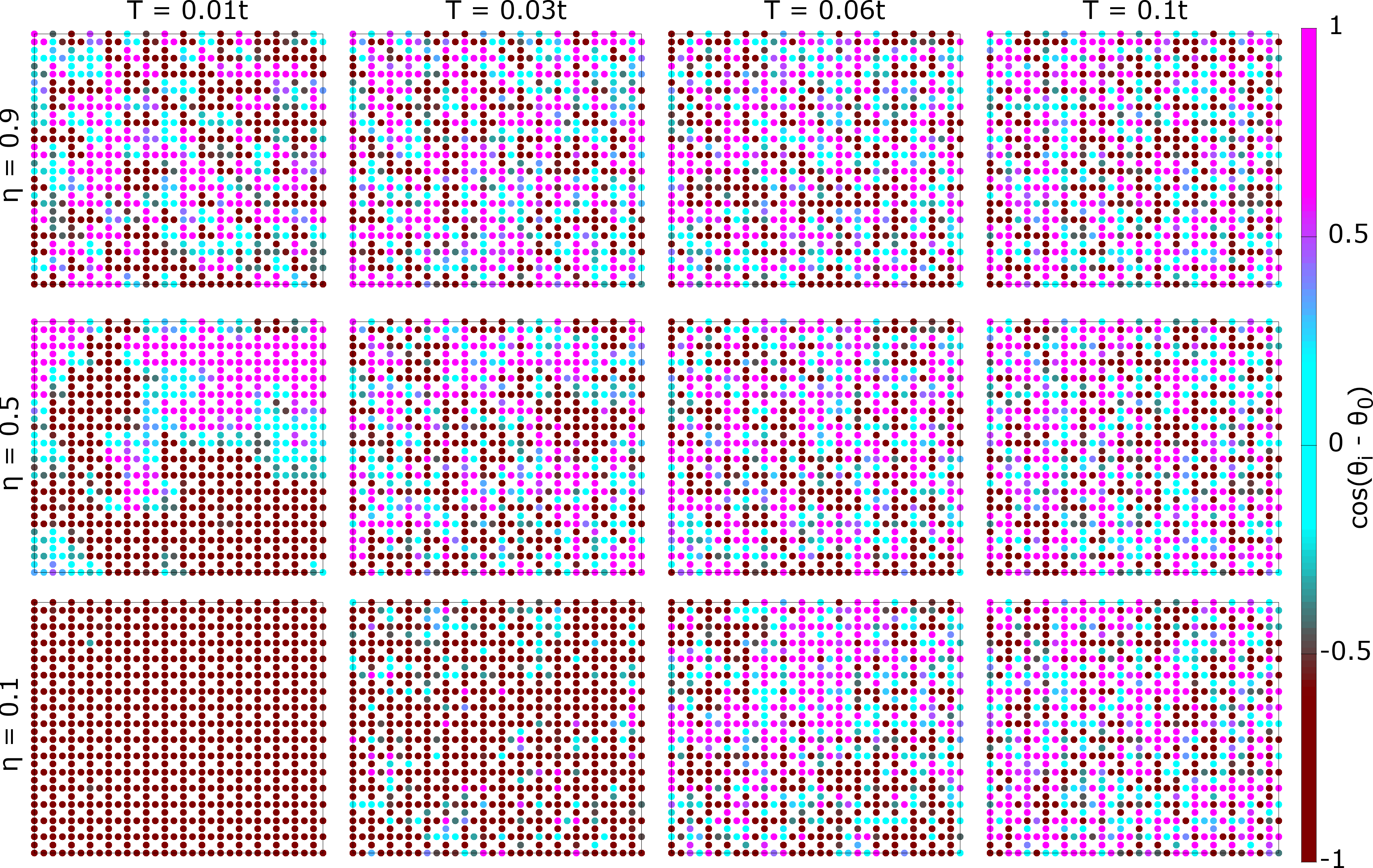}
\caption{\label{fig13} Color online: Real space snapshots of pairing field phase correlation as it evolve with
temperature at $U=2t$ and selected strain cross sections. At weak strain ($\eta=0.1$) the phase 
correlation is long ranged and uniform, which progressively randomizes with temperature. The 
intermediate strain regime of $\eta=0.5$ shows spatial suppression of phase coherence at isolated regimes 
even at the lowest temperature, suggesting weakening of the superconducting order. The large 
strain regime ($\eta=0.9$) is an insulator as suggested by the absence of phase coherence even at 
the lowest temperature. Each point in the map indicates the pairing field phase correlation at that 
particular lattice site with the corresponding weight (magnitude) being given by the color bar.}
\end{figure*}

\subsubsection{Thermal behavior}

We now investigate the fate of the strain induced superconductor-insulator transition 
at finite temperature. For this we present the thermal evolution of the average phase correlation 
at different magnitudes of strain, in Fig. \ref{fig11}(a). As before, the point of inflection corresponds 
to the loss of long range phase coherence, {\it i. e.} T$_{c}$. We note that increasing strain leads 
to progressive suppression of T$_{c}$. The corresponding distribution of the pairing field 
amplitude at selected strain are shown in the panels (b), (c) and (d) of Fig. \ref{fig11}. 
We note that the distribution is bimodal at the lowest temperature for any strain, indicating 
the difference in the contribution of the bond and rim sites towards the superconducting 
pairing. While for the unstrained case we had shown that increasing interaction tends to 
homogenize the contribution from the bond and rim sites, Fig. \ref{fig11} suggests that increasing 
strain leads to a larger difference between the contribution from the bond and rim 
sites. In other words, strain enhances the {\it bipartiteness} of the lattice. We note from 
the distributions that the contribution towards superconducting pairing field amplitude 
from the flat band is $\sim$ 1.5  times of that from the dispersive bands. The simple 
reason for the same is the localization of large number of energy states by the flat band. 
As expected, temperature randomizes the pairing field amplitude and leads to a broader 
distribution. Moreover, the peak amplitude shifts towards larger values in agreement 
with the fact that thermal fluctuations enhance the mean pairing field amplitude. 

\subsubsection{Real space maps}

The real space maps corresponding to the pairing field amplitude and phase correlation shown 
in Fig. \ref{fig12} and \ref{fig13} bear out the information presented through the distributions. 
At the lowest temperature the contribution to the pairing field principally arises from the flat bands 
located at the bond sites, irrespective of the magnitude of strain. Increasing temperature 
randomizes the pairing field amplitude leading to isolated islands of suppressed or enhanced 
pairing. We note that even at large strain ($\eta = 0.9$) the magnitude of $| \Delta_{i}|$'s remain 
fairly robust over a large fraction of the lattice, ensuring a gapped quasiparticle spectra. 

In contrast to the pairing field amplitude, the long range phase coherence is rapidly suppressed 
with strain, as demonstrated in Fig. \ref{fig13}. At weak strain ($\eta=0.1$), the system has long range phase 
coherence at the lowest temperature, which survives upto intermediate temperatures (T $\sim$ T$_{c}$). 
For T$>$T$_{c}$, isolated regions of suppressed phase coherence are realized in the system. 
The phase is akin to the pseudogap regime discussed before, where short range pair correlations 
survive without long range order. Further rise in temperature at this strain randomizes the 
phase completely, leading to complete loss of superconductivity. At intermediate strain 
($\eta = 0.5$) the phase coherence is significantly suppressed even at the lowest temperature. 
Consequently, the corresponding T$_{c}$ is suppressed. Increase in temperature leads to rapid 
loss of phase coherence, thereby killing off superconductivity. Note that at this strain there 
is no pseudogap regime. The large strain ($\eta=0.9$) 
regime lacks any phase coherence even at the lowest temperature. As a result, this regime can be 
broadly thought of to be a insulating state. Thermal fluctuations do not affect the 
state significantly.
The combined picture that emerges from the pairing field amplitude and phase coherence at
large strain is that of a bosonic insulator. Such an insulator is characterized by, 
(i) finite pairing field amplitude, (ii) lack of phase coherence and (iii) large single particle
spectral gap at the Fermi level. 
While (i) and (ii) are shown in Fig. \ref{fig12} and \ref{fig13}, we discuss (iii) 
in the next section pertaining to quasiparticle behavior.    

\begin{figure*}
\includegraphics[height=4.0cm,width=12cm,angle=0]{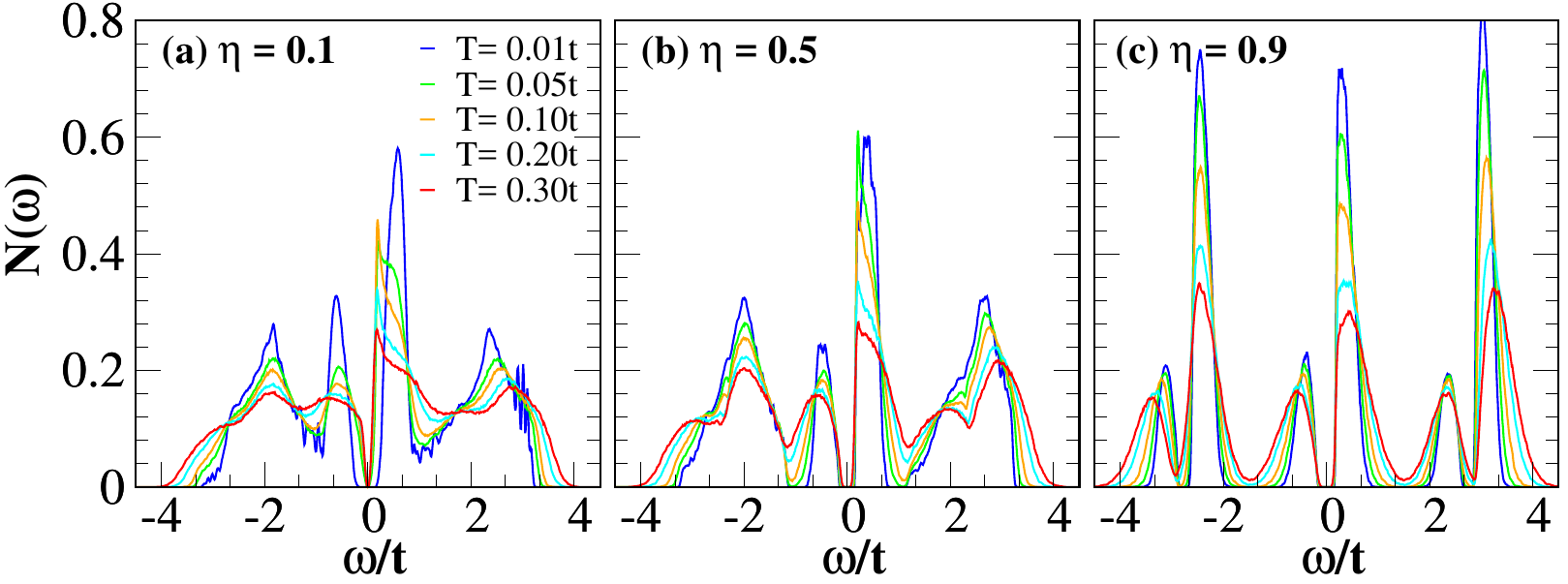}
\hspace{0.3cm}
\includegraphics[height=4.0cm,width=5cm,angle=0]{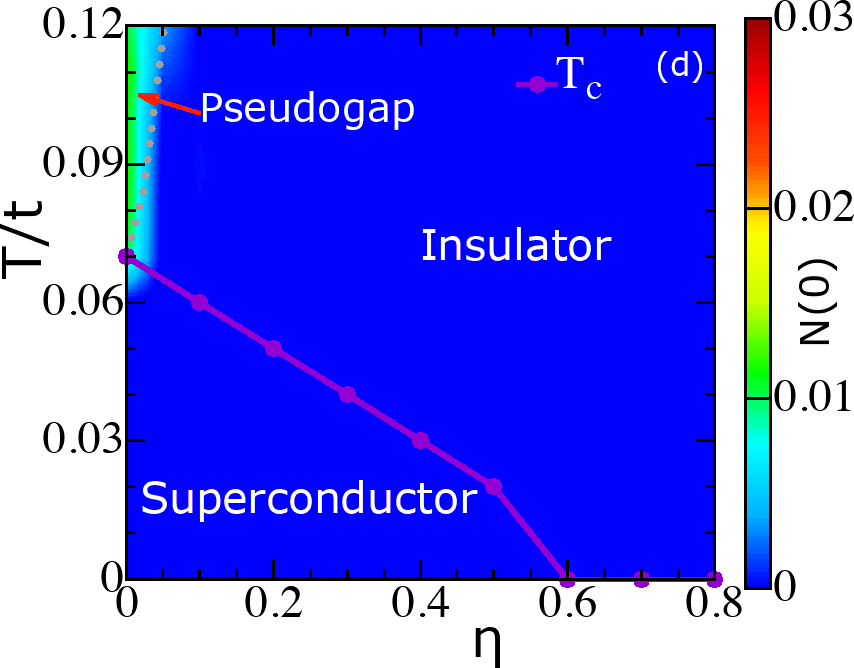}
\caption{\label{fig14} Color online: (a)-(c) Thermal evolution of single particle density of states (DOS) at 
$U=2t$ and selected 
$\eta$ values. At small strain ($\eta=0.1$), the system undergoes a transition 
from a gapped superconductor to a pseudogapped regime with increasing temperature. 
At intermediate ($\eta=0.5$) and strong ($\eta=0.9$) strain, the gap at the Fermi level 
is immune to thermal evolution, owing to the flat band. 
Additionally, away from the Fermi level the DOS gaps out, indicating the localization of the states.
(d) Map corresponding to the single particle DOS at the Fermi level, $N(0)$, across the $\eta-T$ plane, 
at $U=2t$. The magnitude of the spectral weight is indicated by color.}
\end{figure*}
 
\subsubsection{Quasiparticle signatures}

In Fig. \ref{fig14} we show the thermal evolution of single particle density of states (DOS) at selected 
temperature and strain values, at $U=2$t. Fig. \ref{fig14}(a) shows the behavior at weak strain of $\eta=0.1$. 
The low temperature phase corresponds to a gapped BCS-like superconductor 
as suggested by the sharp coherence peaks at the gap edges. 
There are prominent satellite peaks corresponding to the two dispersive bands 
away from the Fermi level. Increasing temperature leads to progressive piling up of 
spectral weight at the Fermi level along with the smearing out of the coherence peaks, 
leading to a weak pseudogap like behavior. The loss of superconductivity in this case is signalled by 
the smearing out of the coherence peaks for $T \ge 0.06$t. In Fig. \ref{fig14}(b) we show the thermal 
evolution of the DOS at an intermediate strain ($\eta = 0.5$). The lowest temperature state at 
this parameter regime is a (BEC-like) superconductor with a finite gap at the Fermi level. 
Note that the gap at the Fermi level is nearly immune to thermal fluctuations. The satellite peaks away
from the Fermi level gets progressively robust with increasing strain, as the energy states get localized. 
At large strain, $\eta = 0.9$, the unit cells get nearly decoupled from each other  (see Fig. \ref{fig14}(c)). 
The dispersive bands flatten out and the spectra now comprises of three flat bands 
(each split into two) with localized energy states. Long range superconducting order is lost in this regime 
and the system behaves as a bosonic insulator.

In Fig. \ref{fig14}(d), we show the single particle DOS at the Fermi level, $N(0)$, at $U$=2t, 
in the $\eta-T$ plane. While the flat band leads to a finite gap across most of the $\eta-T$ plane, 
in the regime of weak strain, a small window of pseudogap phase emerges at high temperatures, as 
signalled by the non zero spectral weight at the Fermi level.
The strain induced SIT discussed in this paper is generic for $x-y$ anisotropy 
in the Lieb lattice. The $N(0)$ map shows that over a large part of the parameter space 
the system undergoes transition from a bosonic superconductor to a bosonic insulator.

\section{Discussion and conclusion}

In this paper we have shown a superconductor-insulator quantum phase transition, 
driven by strain. Across a critical strain $\eta_{c}$, we have characterized this SIT 
based on thermodynamic signatures. 
We have compared and contrasted our inferences {\it vis a vis} with those 
drawn based on the behavior of superfluid weight \cite{torma2017_prl}. 
Our results suggest that {\it the loss of superconducting order with 
increasing strain is dictated by the loss of long range phase coherence even though 
local pair correlations survive. We have further demonstrated a strain induced BCS-BEC like  
crossover in the regime of weak interaction, in this system.} Moreover, in agreement with 
the inference drawn based on the behavior of superfluid weight, our results show that 
the bond sites of the lattice gives larger contribution to superconducting pairing as 
compared to that of the rim sites. Applying strain is a novel but 
certainly not an unique route to realize SIT. A more conventional route is disorder 
driven SIT. While the final outcome is similar through both the approaches, the underlying 
mechanisms are very different. We touch upon them below.

\subsection{Disorder vs strain induced SIT}

Disorder (as introduced via the randomness of chemical potential in the system) 
induced SIT is a well studied subject \cite{ghoshal1998,ghoshal2001,bouadim2011,trivedi1996,trivedi1999}.
The fundamental observation in case of disorder induced SIT is the fragmentation of the superconducting state
(both in superconducting amplitude and phase correlation) as a function of increasing
disorder. The resulting ``insulating'' state is basically a non trivial state with
localized pairs. The single particle spectra at the Fermi level remains 
gapped, in spite of the Griffith's effects of the rare regions \cite{ghoshal1998,ghoshal2001,bouadim2011}. 
The two particle gap $\omega_{pair}$, on the other hand is not a hard gap \cite{bouadim2011,trivedi2014}.

On the other hand for a strain induced (disorder free) SIT, there is no spatial inhomogeneity
in the superconducting amplitude across the transition. Both the superconductor and insulator
phases are characterized by large and homogeneous pairing field amplitude corresponding
to a BEC-like state and a bosonic insulator, respectively. The system discussed in this manuscript falls
in this category as is made evident through the spatial maps of pairing field amplitude
and phase correlation. The single particle spectra is hard gapped across the phase
transition. Discussion of two particle gap is beyond the scope of this work, however,
existing literature on disorder free transition suggests that the two-particle gap $\omega_{pair}$
is hard gapped across such SIT \cite{trivedi2016}. Investigation of the 
effect of flat bands on the two-particle gap is a subject worth pursuing in future.

\subsection{Connection with experiments}

In the context of solid state materials the survival of local pair 
correlations can be probed through STM measurements and tunnelling conductance maps. 
Recently, it has been found that two-dimensional (2D) organic system 
of $sp^{2}$-carbon-conjugated covalent organic framework ($sp^{2}c$-COF) is a material 
realization of the Lieb lattice \cite{jin_science2017}. Based on the experimental 
observation \cite{jin_science2017} and numerical simulations \cite{liu_natcom2019}
it was found that the material has a Lieb lattice like structure with staggered hopping 
(strain), similar to the situation discussed in the present work. $sp^{2}c$-COF 
has been experimentally observed to exhibit metal insulator transition as well as 
unconventional magnetic instability, which were attributed to the strained lattice structure 
of this material \cite{jin_science2017}. Such a discovery of material realization of Lieb-like lattice indeed 
opens up scope for future exploration for systems exhibiting novel phase transitions and 
unconventional quantum phases. One such ``yet to be discovered'' possibility is the SIT 
discussed in the present paper.  
 
While the application of strain through staggered hopping is an experimentally challenging task 
for solid state materials, it is certainly more feasible for ultracold atomic gases and photonic lattices, 
where the control parameters are tunable. As on today, one of the biggest 
challenges faced by the ultracold atomic gas experiments is that sufficiently low temperatures 
could not be attained so as to realize superfluid ground state. However, capturing the signatures 
of pair correlations at high temperatures is certainly an achievable goal for these experiments 
and has already been done for Fermi gases \cite{chin2004,gaebler2010}, 
population imbalanced ultracold atomic gases \cite{ketterle_nature2008,ketterle_science2007} etc. 
The spectroscopic probes to capture the signature of preformed pairs in these systems are 
radio frequency (rf) spectroscopy, momentum resolved rf spectroscopy etc 
\cite{gaebler2010,ketterle_nature2008,ketterle_science2007,chin2004,zoller2000,zoller2001,levin2005,torma2004,jin2003,ketterle2003,ketterle2003_prl,strinati2008,strinati2009,grimm2011,ketterle2008_prl,stewart2008,torma2004_prl}. 
We believe that similar techniques can be utilized to understand the physics of superfluid pair 
formation in ultracold atomic gases on Lieb lattice. The finite temperature results pertaining to 
the quasiparticle signatures discussed in this paper are expected to provide useful benchmarks 
for the experiments based on spectroscopic measurements. 

Finally, ours is not the first system to show disorder free BEC-bosonic
insulator SIT at weak coupling \cite{trivedi2016}. However, to the best of our knowledge, 
we have demonstrated the same for the first time for a flat band system. 
Moreover, not only this system has been realized as an artificial designer lattice, 
recently a solid state material with similar lattice architecture has been realized. 
This makes our predictions easily verifiable in experimental settings.

The SIT discussed in this manuscript is not a consequence 
of the flat band in the system. It is the band structure reconstruction (change in topology of the bands)
due to the applied strain which gives rise to the SIT.
The spectral and real space signatures show that there are bosonic 
phases on the either side of the SIT (BEC and bosonic insulator), even in the weak 
coupling regime. Our results suggest that the nature of the superconducting pairing field 
as contributed by the flat and dispersive bands might be different. While the contribution 
of spectral weight by the flat band is always bosonic, the dispersive band contributions 
depend on the applied strain and/or interactions, which facilitates the BCS-BEC crossover 
like physics.
Our analysis further shows that the exclusion of phase fluctuations 
of the pairing field in the calculation of this SIT would lead to an incorrect phase diagram.
Since the pairing field amplitude remains robust across the transition, mapping
out the phase diagram based on it would show the survival of superconducting
order even for an arbitrarily large strain.

In conclusion, in this work we have established the strain induced superconductor-insulator
quantum phase transition on the Lieb lattice. Based on thermodynamic signatures 
we have demonstrated how global superconductivity is destroyed with loss of long range phase coherence 
and also presented a mean field estimate of critical strain across which 
the quantum phase transition takes place between a bosonic superconductor and a bosonic insulator. 
Further, we have demonstrated that strain alters the characteristic of the underlying 
superconducting order and leads to a strain induced BCS-BEC crossover.
This work is the first demonstration of disorder free SIT 
on the Lieb lattice and is expected to open up experimental avenues both for the solid state 
and ultracold atomic gas communities. Further, we have discussed several indicators 
which we believe should be accessible through the existing experimental probes. 
Apart from being engineered in the artificial systems such as photonic lattices or ultracold atomic gases, 
the Lieb lattice has been found to be the building blocks of solid state materials 
such as cuprates, 2D organic materials etc. Such materials require a more complex model 
than the one discussed in this paper. While it is non trivial to include all the interactions and 
parameters relevant for such materials in a tractable theoretical model, 
the future works aim towards capturing the physics of these materials through more realistic models. 

\appendix

\section{Superconducting gap}

\begin{figure}[b]
\includegraphics[height=5.3cm,width=8.5cm,angle=0]{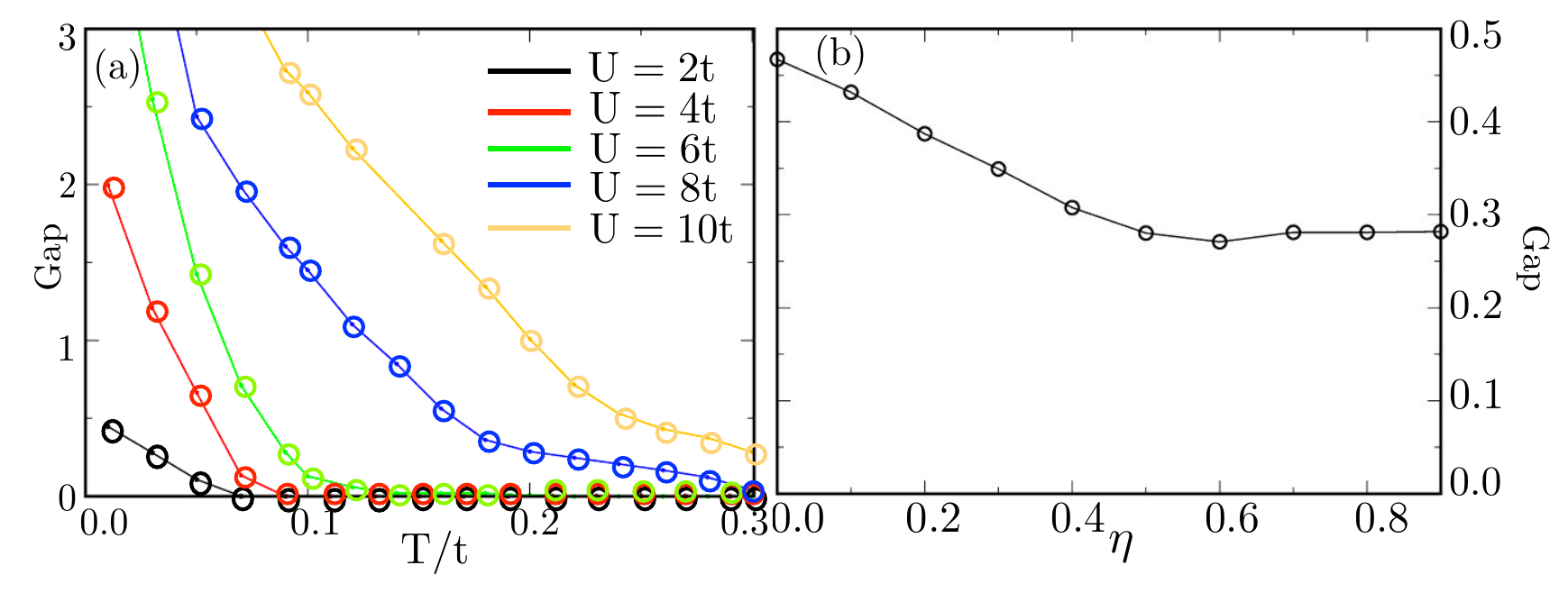}
\caption{\label{fig15} Color online: (a) Thermal evolution of quasiparticle spectral gap at selected
interactions, for an unstrained Lieb lattice. (b) Evolution of spectral gap as a function
of strain across the quantum phase transition at $\eta_{c}$=0.6 and $U=2t$.}
\end{figure}

In the earlier sections we have discussed that the loss of superconductivity is 
dictated by the loss of long range phase coherence, 
even though the pairing field amplitude remains finite. While a mean field treatment is 
expected to overestimate the robustness of the ordered state, a recent comparison between 
the mean field and DMFT results suggest that in the limit of weak coupling and low temperatures 
the agreement between the two approaches is reasonable \cite{torma2017_prl}. At higher temperatures 
the DMFT treatment expectedly fares better since it takes into account some of the spatial fluctuations.
Stronger interactions tend to make the approach less accurate (for $d < \infty$) as spatial 
fluctuations dominate the high temperature scenario. The characterization of the superconducting 
state in ref.(\cite{torma2017_prl}) is carried out based on the local (sublattice) superconducting 
``order parameter'',  determined by solving the self consistent gap equation. 
In our present numerical scheme the global 
superconducting gap can be determined from the single particle DOS at the Fermi level and we 
show the same in Fig. \ref{fig15}. Panel (a) of Fig. \ref{fig15} corresponds to the thermal evolution of the 
superconducting gap at different interaction strengths corresponding to the weak, intermediate and 
strong coupling regimes. We note that in the regime of weak interactions the behavior of the gap 
is in agreement with the one obtained from DMFT calculations with the gap closing at the T$_{c}$ 
\cite{torma2017_prl}. At stronger interactions however the gap is robust even at higher temperatures.
We emphasize that the high temperature gap correspond to a correlated bosonic insulator rather than 
a superconductor. In other words, away from the weak coupling regime the superconducting order 
is no longer tied to the quasiparticle spectral gap at the Fermi level indicating the breakdown of the mean 
field theory.  Though the results corresponding to DMFT analysis in ref.(\cite{torma2017_prl}) pertains
to the weak interaction regime only, we believe that neglect of spatial fluctuations would lead 
to incorrect thermal scales away from the weak coupling within the framework of DMFT. 
  
Fig. \ref{fig15}(b) shows the evolution of quasiparticle spectral gap at the Fermi level 
with strain at the ground state and $U$=2t. For $\eta \le \eta_{c}$ there is monotonic decrease in the gap 
with increasing strain since the 
superconducting correlations weaken as the unit cells progressively decouple from each other with increasing 
strain. For strain $\eta > \eta_{c}$ the system is in an insulating state and the corresponding 
spectral gap at the Fermi level is immune to the effect of strain. 
The change in the size of the gap across the SIT is $\sim$ 18$\%$.
Note that the spectral gap does 
not vanish with increasing strain, rather it is the phase correlation which undergoes transition across the quantum 
critical point.

\section{Finite system size effect}

\begin{figure}
\includegraphics[height=5.3cm,width=8.5cm,angle=0]{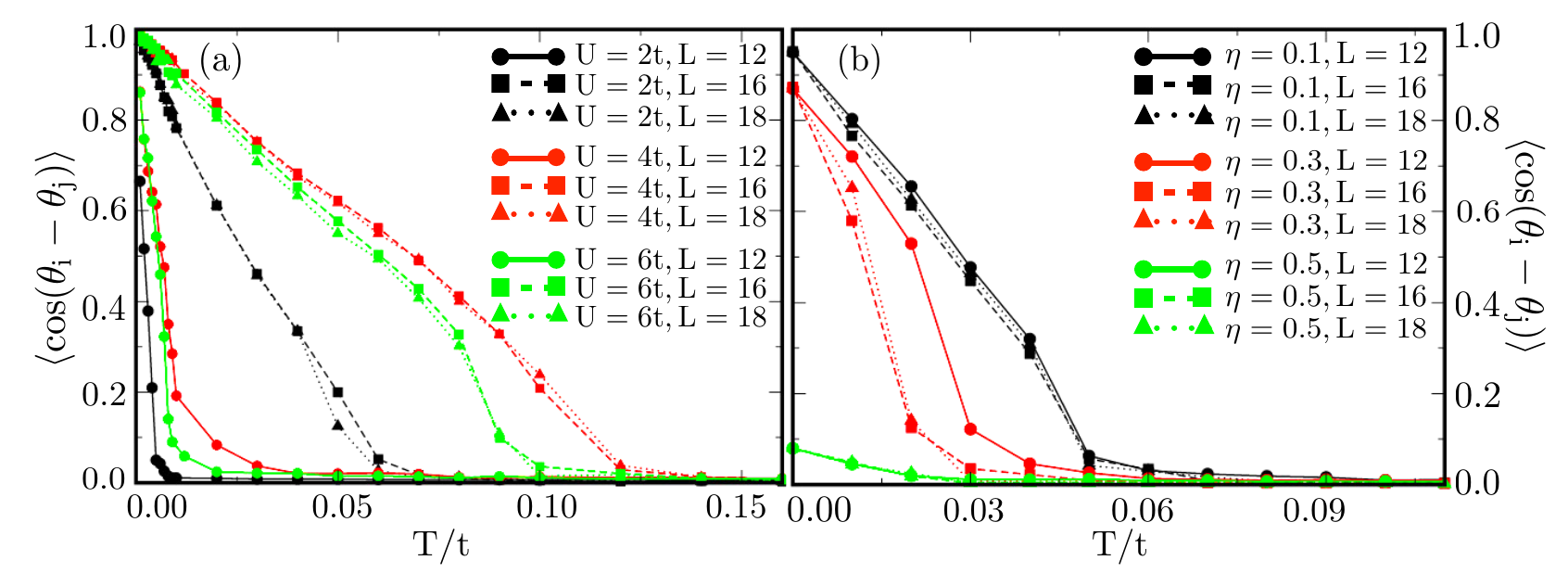}
\caption{\label{fig16} Color online: (a) Thermal evolution of average phase correlation at selected 
interactions of $U=2t, 4t, 6t$ and system sizes of $L=12$, $L=16$ and $L=18$. 
(b) Thermal evolution of average phase correlation at $U=2t$ and selected strain values
of $\eta=0.1$, $\eta=0.3$ and $\eta=0.5$ for the system sizes same as in (a).}
\end{figure}

The results discussed in this paper correspond to a system size of $N=768$ ($L=16$). 
Any lattice simulation is however susceptible to finite size effects and in order to 
verify whether our results are robust against the choice of the system size we have 
carried out the numerical simulations at different system sizes. In Fig. \ref{fig16} we show 
the thermal evolution of superconducting phase correlation at, (a) selected interactions 
$U$=2t, 4t and 6t (for the unstrained case) and (b) selected strain values $\eta$=0.1, 0.3 
and 0.5 (for the strained case) at three different lattice sizes of $N=432 (L=12)$, 
$N=768 (L=16)$ and $N=972 (L=18)$.
We note that while there are noticeable finite size effect 
at $N=432$ as observed via the underestimation of the T$_{c}$ scales, the larger 
system sizes $N \ge 768$ are immune to finite size effects, suggesting that the 
results discussed in this paper are fairly robust against the choice of the system 
sizes. Our finite system size analysis further justifies that the strongly suppressed 
T$_{c}$ at strong coupling ($U$=8t) as observed in DQMC study \cite{scalletar2014} is 
an artifact of small system size.  

\section{Fixed chemical potential vs fixed number density}

\begin{figure*}
\includegraphics[height=6.0cm,width=16cm,angle=0]{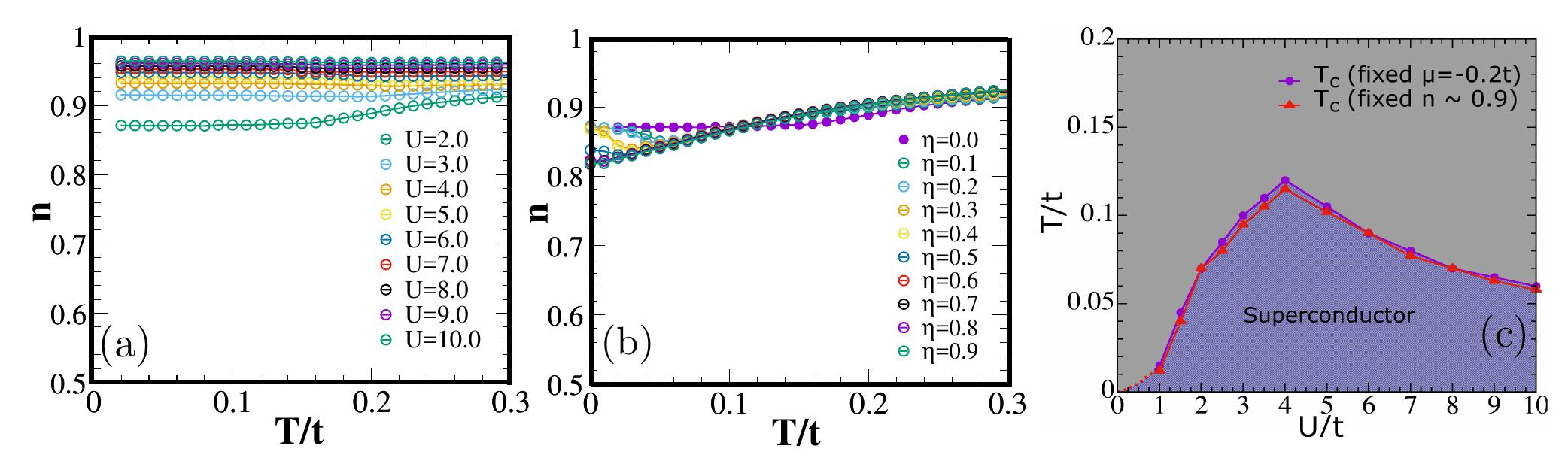}
\caption{\label{fig17} Color online: Thermal evolution of fermionic number densities at
different (a) interactions and (b) strain. (c) BCS-BEC crossover at $\eta=0.0$
as computed at a fixed chemical potential ($\mu=-0.2t$) and a fixed number density ($n\approx 0.9$).}
\end{figure*}

Our calculations are carried out in grand canonical ensemble with a 
fixed chemical potential of $\mu = -0.2t$. In many situations such as inclusion of disorder in the 
system, the fermionic number density varies significantly with the disorder strength/concentration 
in calculations carried out in grand canonical ensemble. The inferences that one would draw 
from such a calculation will be significantly different from those obtained from the calculations 
carried out at fixed number density. In order to understand the dependence of the number density 
on temperature, interaction and strain at $\mu = -0.2t$, we show the same in Fig. \ref{fig17}. 
The choice of $\mu$=-0.2t corresponds to a fermionic
filling of $n \approx 0.9$ and for $|U| \ge 2t$ remains almost independent
of the choice of the interaction strength and temperature. Importantly, in the
absence of any competing order at the low temperature (as in the present model),
we do not expect a small drift in the fermionic number density to alter the state of the
system significantly. We have verified the same across the BCS-BEC crossover and
the results are shown in Fig. \ref{fig17}(c). The figure shows the comparison of the BCS-BEC
crossover as calculated at a fixed chemical potential of $\mu=-0.2t$ and at a fixed fermionic
number density of $n\approx 0.9$. The results suggest that for the system (model) and parameter regime
under consideration, calculations carried out in grand canonical ensemble captures
the behavior of the system accurately.

\section{Static path approximation (SPA) in comparison with other techniques}
The Hubbard model at intermediate interaction requires a non perturbative
solution. The exponential growth in the dimension of the Hilbert
space rules out the use of exact diagonalization except for very
small system sizes. The ``exact'' tool of choice is quantum Monte Carlo (QMC)
and all approximations are bench marked against it. While QMC can be
implemented via various approaches, the method below easiest reveals
the connection to our approach \cite{evenson1970, dubi2007, karmakar2016}:

The Hubbard partition function is written as a functional integral
over Grassmann fields $\psi_{i \sigma} (\tau), {\bar \psi}_{i \sigma}(\tau)$ as,

\begin{eqnarray}
Z & = &  \int {\cal D} \psi {\cal D} {\bar \psi}e^{-S[ \psi,  {\bar \psi}]} \cr
S & = & \int_0^{\beta} d\tau[\sum_{ij,\sigma, \sigma^{\prime}} \{ {\bar \psi}_{i\sigma} 
((\partial_{\tau}-\mu)\delta_{ij}  - t_{ij})
\psi_{j\sigma} \} \nonumber \\ && + \vert U \vert \sum_{\langle ij \rangle, \sigma, \sigma^{\prime}} 
{\bar \psi}_{i\sigma} {\psi}_{i\sigma}
{\bar \psi}_{j\sigma^{\prime}} {\psi}_{j\sigma^{\prime}}]
\end{eqnarray} 

Only quadratic path integrals can be exactly evaluated. Since
the interaction generates a quartic term in the $\psi$'s the
partition function cannot be immediately evaluated.

The quartic term is ``decoupled'' exactly through a Hubbard-Stratonovich
transformation in terms of pairing fields $\Delta_{i}(\tau),{\bar                                                      
\Delta}_{i}(\tau)$. This induces a term $\Delta_{i}\bar \psi_{i\uparrow}(\tau)\bar \psi_{i\downarrow}(\tau)$ 
in the action,

\begin{eqnarray}
Z & = & \int  {\cal D} \Delta {\cal D} \Delta^* {\cal D} \psi {\cal D} {\bar \psi}
e^{ - S_1[ \psi,  {\bar \psi}, \Delta, \Delta^*]} \cr S_1 & = & \int_0^{\beta} d\tau[
\sum_{ij,\sigma} \{{\bar \psi}_{i\sigma} ((\partial_{\tau} - \mu)\delta_{ij}  - t_{ij})
\psi_{j\sigma}\}  \nonumber \\ && + \sum_{i} \{\Delta_{i}(\tau) {\bar \psi}_{i \uparrow}(\tau)
{\bar \psi}_{i \downarrow}(\tau)  + h. c. + {\vert \Delta_{i}\vert^2 \over \vert U \vert}\}]
\end{eqnarray} 

The $\psi$ integral is now quadratic but at the cost of an
additional integration over the fields $\Delta_{i}(\tau)$ and
$\Delta_{i}^{*}(\tau)$.
The ``weight factor'' for the  $\Delta_{i}$ configurations can be determined by
integrating out the  $\psi,{\bar \psi}$, and using these weighted
configurations one goes back and computes fermionic properties.
Formally,

\begin{eqnarray}
Z &=&  \int  {\cal D} \Delta {\cal D} \Delta^*e^{-S_2 [ \Delta, \Delta^*]} \cr
S_2 & = & \log[Det[{\cal G}^{-1} - \Delta]] + {\vert \Delta_{i}\vert^2 \over \vert U \vert}
\end{eqnarray} 
where, ${\cal G}$ is the electron Green's function in a $\{\Delta_{i}\}$ background.

The weight factor for an arbitrary space-time configuration $\Delta_{i}(\tau)$
involves computation of the fermionic determinant in that background.
If we write the auxiliary field $\Delta_{i}(\tau)$ in terms of its Matsubara
modes, as $\Delta_{i}(\Omega_n)$, then the various approximations can be
readily recognized and compared.

\begin{itemize}
\item{Quantum Monte Carlo retains the full ``$i,\Omega_n$'' dependence of $\Delta$
computing $\log[Det[{\cal G}^{-1} - \Delta]]$ iteratively for importance
sampling. The approach is valid at all $T$, but does not readily yield
real frequency spectra.}
\item{Mean field theory (MFT) is time independent, neglects the
phase fluctuations completely but can handle spatial inhomogeneity
in amplitude of the pairing field. Thus, $\Delta_{i}(i\Omega_n) \rightarrow \Delta_{i}$.
When the MF order parameter vanishes at high temperature the theory trivializes.}
\item{Our static path approximation (SPA) approach retains the full spatial dependence
in $\Delta$ but keeps only the $\Omega_n=0$ mode,
{\it i.e}, $\Delta_{i}(\Omega_n) \rightarrow \Delta_{i}$.
It thus includes classical fluctuations of arbitrary magnitude but no quantum ($\Omega_n \neq 0$)
fluctuations. One may
consider different temperature regimes: \\
(1)~$T=0$: since classical
fluctuations die off at $T=0$, SPA reduces to standard Bogoliubov-de Gennes (BdG)
MFT. \\
(2)~At $T \neq 0$ we consider not
just the saddle point configuration but {\it all configurations} following
the weight $e^{-S_2}$ above. These involve the classical amplitude and
phase fluctuations of the order parameter, and the BdG equations are solved
{\it in all these configurations} to compute the thermally averaged
properties. This approach suppresses the order much
quicker than in MFT. \\
(3)~High $T$: since the $\Omega_n=0$ mode
dominates the exact partition function the SPA approach
becomes exact as $T \rightarrow \infty$.}
\item{DMFT: for completeness we mention that DMFT retains the full dynamics
but keeps $\Delta$ at  effectively one site, {\it i.e},
$\Delta_{i}(\Omega_n) \rightarrow \Delta(\Omega_n)$.
This is exact when dimensionality $D \rightarrow \infty$.}
\end{itemize}

\section{Effect of quantum fluctuations}

The static path approximation used in this work gets progressively accurate with
increasing temperature and is akin to the mean field theory at the ground state.
To that extent the ground state phase diagram as reported in this manuscript is 
a mean field estimate, where the quantum fluctuations are being
neglected. We argue that the neglect of quantum fluctuations is a reasonable
approximation for the present problem where the only gapless mode are the
$XY$-type low energy excitations of the superconducting phase. Models with
$XY$ symmetry are well known to have long range order in two dimensions and a BKT transition
at finite temperature. The issue of fluctuations thus reduces to verifying
how well the $U(1)$ superconducting T$_{c}$ is captured by our approach as
compared to that obtained via QMC. While QMC results for the present problem
(strain induced SIT) are unavailable for comparison, the agreement between the results obtained
through QMC and SPA for other systems (e. g. attractive Hubbard model on 2D
square lattice \cite{tarat_epjb}) suggests that the relevant fluctuations are suitably captured
by our numerical technique.

Further, as we have discussed in our manuscript, in a recent work on
attractive Hubbard model on the Lieb lattice the results obtained via
the mean field theory (MFT) has been compared with those
obtained via DMFT (which takes into account quantum fluctuations) \cite{torma2017_prl}.
It has been demonstrated that at the ground state the results obtained
by both the techniques are in fairly good agreement with each other,
suggesting that the mean field approach to the problem is good enough
to capture the ground state. At high temperatures
MFT expectedly overestimates the thermal scales.

Based on the above discussion we emphasize that MFT is a reasonably good approximation
to capture the ground state physics of the system discussed in this manuscript, and we
do not expect any qualitative changes in the same through inclusion of quantum fluctuations.

\begin{acknowledgments}
N. S. acknowledges use of the NSCC Singapore ASPIRE-1 cluster for some part of the numerical simulations.
M. K. acknowledges use of the HPC clusters at HRI, India for detailed numerical calculations.
\end{acknowledgments}
 
\bibliographystyle{apsrev4-1}
\bibliography{lieb}
\end{document}